\documentclass[12pt]{article}
\usepackage{graphics}
\usepackage[dvips]{graphicx}
\usepackage{mathrsfs}
\usepackage{amssymb}
\usepackage{verbatim}
\usepackage{float}
\newcommand{\be}{\begin{equation}}
\newcommand{\ee}{\end{equation}}
\newcommand{\bea}{\begin{eqnarray}}
\newcommand{\eea}{\end{eqnarray}}
\newcommand{\nnmb}{\nonumber}

\def\4vol{{\int d^4x \sqrt{-g}}}

\newcommand{\nc}{\newcommand}

\nc{\nt}{\tilde{N}}
\nc{\ra}{\rightarrow}
\nc{\lsim}{\begin{array}{c}\,\sim\vspace{-21pt}\\< \end{array}}
\nc{\gsim}{\begin{array}{c}\sim\vspace{-21pt}\\> \end{array}}
\nc{\tnt}{\tilde{N}}
\nc{\tst}{\tilde{t}}
\nc{\beq}{\begin{equation}}
\nc{\eeq}{\end{equation}}
\nc{\gev}{\mbox{GeV}}
\nc{\tev}{\mbox{TeV}}
\nc{\eev}{\mbox{eV}}
\nc{\db}{\Delta b}
\nc{\met}{\,/\!\!\!\!E_T}
\nc{\mpt}{\,/\!\!\!p_T}
\nc{\nfive}{N_{{\bf 5}\oplus\bar{\bf 5}}}
\nc{\five}{{\bf 5}\oplus\bar{\bf 5}}

\nc{\LL}{L}
\nc{\vv}{\tilde{v}}
\nc{\bit}{\begin{itemize}}
\nc{\eit}{\end{itemize}}

\begin{document}

\begin{titlepage}
\renewcommand{\thefootnote}{\fnsymbol{footnote}}
\noindent
\begin{flushright}
MCTP-06-37 \\
SU-4252-842 
\end{flushright}

\vspace{1cm}

\begin{center}
  \begin{Large}
    \begin{bf}
Connecting (Supersymmetry) LHC Measurements with High Scale Theories
     \end{bf}
  \end{Large}
\end{center}
\vspace{0.2cm}
\begin{center}
\begin{large}
Gordon~L.~Kane$^{(a)}$\footnote{gkane@umich.edu},
Piyush~Kumar$^{(a)}$\footnote{kpiyush@umich.edu},
David~E.~Morrissey$^{(a)}$\footnote{dmorri@umich.edu},
Manuel~Toharia$^{(a,b)}$\footnote{mtoharia@physics.syr.edu}\\
\end{large}
  \vspace{0.3cm}
  \begin{it}
$^{(a)}$ Michigan Center for Theoretical Physics (MCTP) \\
Department of Physics, University of Michigan, Ann Arbor, MI 48109\\
\vspace{0.2cm}
$^{(b)}$ Department of Physics, Syracuse University, Syracuse, NY, 13244-1130

\vspace{0.1cm}
\end{it}

\end{center}

\center{\today}

\renewcommand{\thefootnote}{\arabic{footnote}}
\setcounter{footnote}{0}

\begin{abstract}

  If supersymmetry is discovered at the LHC, the measured spectrum of 
superpartner masses and couplings will allow us to probe the origins 
of supersymmetry breaking.  However, to connect the collider-scale
Lagrangian soft parameters to the more fundamental theory from which 
they arise, it is usually necessary to evolve them to higher scales.  
The apparent unification 
of gauge couplings restricts the possible forms of new physics above
the electroweak scale, and suggests that such an extrapolation is possible.
Even so, this task is complicated if the low scale
spectrum is not measured completely or precisely, or if there is
new physics at heavy scales beyond the reach of collider experiments.
In this work we study some of these obstacles to running up,
and we investigate how to get around them.  Our main conclusion is 
that even though such obstacles can make it very difficult to accurately
determine the values of all the soft parameters at the high scale,
there exist a number of special combinations of the soft parameters
that can avoid these difficulties.  We also present a systematic application
of our techniques in an explicit example.

\end{abstract}

\vspace{1cm}

\end{titlepage}

\setcounter{page}{2}


\section{Introduction}
 
   It is a remarkable feature of the minimal supersymmetric extension
of the standard model~(MSSM) that the $SU(3)_c$, $SU(2)_L$, and $U(1)_Y$ 
gauge couplings unify at a very high scale, 
of order $10^{16}$~GeV~\cite{Martin:1997ns,Dimopoulos:1981zb}.  
Furthermore, the matter fields of the MSSM, with the exception of 
the pair of Higgs doublets, have precisely the quantum numbers of 
three sets of $\overline{\bf 5}\oplus{\bf 10}$ representations of $SU(5)$.  
These properties may be purely accidental,
but they do suggest a more symmetric unified structure 
at energies only slightly below the Planck scale.  They also offer
a tantalizing hint of the structure of Nature at scales well above
what we will be able to probe directly with colliders such as the LHC.

  Taking these features as being more than accidental, we obtain  
significant constraints on the types of new physics that can arise 
between the electroweak and the grand unification~(GUT) scales.  
Any new phenomenon that enters the effective theory in this energy range
ought to maintain the unification of couplings, and should be consistent
with a (possibly generalized) GUT interpretation.  The simplest scenario 
is a \emph{grand desert}, in which there is essentially no new physics 
at all below the unification scale $M_{GUT}$.  In this case, 
if supersymmetry is discovered at the Tevatron or the LHC, 
it will be possible to extrapolate the measured soft supersymmetry 
breaking parameters to much higher scales using the renormalization group~(RG).
Doing so may help to reveal the details of supersymmetry breaking, 
and possibly also the fundamental theory underlying it.

  If supersymmetry is observed in a collider experiment,
it will be challenging to extract all the supersymmetry breaking parameters
from the collider signals.  While some work has been put into
solving this problem~\cite{Zerwas:2002as}, there is 
still a great deal more that needs to be done.  
The parameters extracted in this way will be subject to 
experimental uncertainties, especially if the supersymmetric spectrum 
is relatively heavy.  There will also be theoretical uncertainties 
from higher loop corrections in relating the physical masses to their
running values~\cite{Pierce:1996zz}.
These uncertainties in the supersymmetry breaking parameters, 
as well as those in the supersymmetric parameters, will complicate 
the extrapolation of the soft masses to high 
energies~\cite{Blair:2000gy,Martin:2001zw}.  
Much of the previous work along these lines has focused on running 
the parameters of particular models from the high scale down.  
This is useful only if the new physics found resembles one of the 
examples studied.  Our goal is to study the running from 
low to high~\cite{Carena:1996km}.

  Evolving the soft parameters from collider energies up 
to much higher scales can also be complicated by new
physics at intermediate energies below $M_{GUT}$.  
The apparent unification of gauge couplings suggests that
if this new physics is charged under the MSSM gauge group,
it should come in the form of complete GUT multiplets or gauge singlets.
Indeed, the observation of very small neutrino masses already 
suggests the existence of new physics in the form of 
very heavy gauge singlet neutrinos~\cite{Mohapatra:2005wg}. 
With new physics that is much heavier than the electroweak scale,
it is often very difficult to study it experimentally, 
or to even deduce its existence.  If we extrapolate the MSSM parameters
without including the effects of heavy new physics, we will obtain
misleading and incorrect values for the high scale values of these
parameters~\cite{Baer:2000gf}.

  In the present work we study some of these potential obstacles 
to the RG evolution of the MSSM soft parameters.  In Section~\ref{sterm} 
we investigate how uncertainties in the low-scale parameter values 
can drastically modify the extrapolated high-scale values.
We focus on the so-called $S$ term (\emph{a.k.a.} the hypercharge $D$ term) 
within the MSSM, which depends on all the soft masses in the theory,
and can have a particularly large effect on the running if some of
these soft masses go unmeasured at the LHC.
In Sections~\ref{gutmult} and \ref{neut}, we study two possible examples
of heavy new physics.  Section~\ref{gutmult} investigates the 
effects of adding complete vector-like GUT multiplets on the running 
of the soft parameters.  Section~\ref{neut} describes how including
heavy Majorana neutrinos to generate small neutrino masses can alter
the running of the MSSM soft parameters.  In Section~\ref{all} we combine
our findings and illustrate how they may be put to use with an
explicit example.
Finally, Section~\ref{conc} is reserved for our conclusions.
A summary of some useful combinations of scalar soft masses is given 
in an Appendix.

  Our main result is that the high scale values of many Lagrangian 
parameters can be very sensitive to uncertainties in their low-scale values,
or to the presence of heavy new physics.  However, in the cases studied 
we also find that there are particular combinations of the 
Lagrangian parameters that are stable under the RG evolution, 
or that are unaffected by the new physics.
These special parameter combinations are therefore especially useful 
for making a comparison with possible high-scale theories.  

  Throughout our analysis, we simplify the RG equations
by setting all flavor non-diagonal soft terms to zero and keeping
only the (diagonal) Yukawa couplings of the third generation.
Under this approximation, we work to two loop order for the running
of the MSSM parameters, and interface with Suspect~2.3.4~\cite{Djouadi:2002ze}
to compute one-loop threshold corrections at the low scale.  
For concreteness, we take this scale to be $500\,\gev$.
The additional running due to new physics introduced at scales 
much larger than the weak scale is only performed at one-loop.  
We also implicitly assume that the mass scale of the messengers 
that communicate supersymmetry breaking to the visible sector lies 
at or above the GUT scale, $M_{GUT} \simeq 2.5\times 10^{16}\,\gev$.
Even so, our methods and general analysis will also be applicable 
to scenarios that have lighter messenger particles, such as gauge 
mediation~\cite{Dine:1993yw,Giudice:1998bp}.  We also neglect the effects 
of hidden sector running, which can be significant if there 
are interacting states in the hidden sector
significantly lighter than $M_{GUT}$~\cite{Luty:2001jh,Dine:2004dv,
Cohen:2006qc}.  While this manuscript was in preparation,
methods similar to those considered in the present work were proposed 
in Ref.~\cite{Cohen:2006qc} to deal with these additional uncertainties 
in the high scale values of the soft parameters.

  In this work, we focus on low-energy supersymmetric models,
and particular forms of intermediate scale new physics.
Despite this restriction, we expect that our 
general techniques will be applicable to other solutions 
of the gauge hierarchy problem, or to more exotic forms 
of new intermediate scale physics.


\section{Uncertainties Due to the \emph{S} Term\label{sterm}}

  The one-loop renormalization group~(RG) equations of the MSSM soft
scalar masses have the form~\cite{Martin:1993zk}
\be
(16\pi^2)\frac{dm_i^2}{dt} = \tilde{X}_i -\sum_{a=1,2,3}8\,g_a^2C^a_i|M_a|^2
+ \frac{6}{5}g_1^2Y_i\,S,
\label{1:msoft}
\ee
where $t=\ln(Q/M_Z)$, $\tilde{X}_i$ is a function of the soft
squared masses and the trilinear couplings, $M_a$ denotes the $a$-th
gaugino mass, and the $S$ term is given by
\bea
S &=& Tr(Y\,m^2)\label{1:sterm}\\
&=& m_{H_u}^2 - m_{H_d}^2 + tr(m_Q^2-2\,m_U^2+m_E^2 + m_D^2-m_L^2)\nnmb
\eea
where the first trace runs over all hypercharge representations, and the
second runs only over flavors.

  The $S$ term is unusual in that it connects the running of any single
soft mass to the soft masses of every other field with non-zero
hypercharge.  Taking linear combinations of the RG equations for
the soft scalar masses, the one-loop running of $S$ in the MSSM 
is given by
\be
(16\pi^2)\,\frac{dS}{dt} = -2\,b_1\,g_1^2\,S,
\label{1:dels}
\ee
where $b_1 = -33/5$ is the one-loop beta-function coefficient.
Using Eq.~(\ref{1:dels}) in Eq.~(\ref{1:msoft}) and neglecting
the Yukawa-dependent terms $\tilde{X}_i$ (which are expected to be 
small for the first and second generations) 
the effect of $S\neq 0$ is to shift the high scale value 
the soft mass would have had were $S(t_0)=0$ by an amount
\beq
\Delta m_i^2(t) = \frac{Y_i}{Tr(Y^2)}\left[
\frac{g_1^2(t)}{g_1^2(t_0)}- 1\right]\,S(t_0).
\label{1:delm2}
\eeq
The one-loop RG equation for $S$ is homogeneous.  Thus, if $S$ 
vanishes at any one scale, it will vanish at all scales (at one-loop).
In both minimal supergravity~(mSUGRA) and 
simple gauge-mediated models, $S$ does
indeed vanish at the (high) input scale, and for this reason
the effects of this term are often ignored. 

   From the low-energy perspective, there is no reason 
for $S(t_0)$ to vanish, and in many cases its effects can be 
extremely important.  Since $g_1$ grows with increasing energy, 
the mass shift due to the $S$ term grows as well.  
For $t_0 \simeq t_{M_Z}$ and $t\to t_{GUT}$, the prefactor in 
Eq.~(\ref{1:delm2}) is about $(0.13)\,Y_i$.  The value of 
$S(t_0)$ depends on all the scalar soft masses, and the experimental 
uncertainty in its value will therefore be set by the least well-measured
scalar mass.  In particular, if one or more of the soft masses aren't
measured at all, $S(t_0)$ is unbounded other than by considerations of
naturalness.  Fortunately, this uncertainty only affects the soft 
scalar masses.  The $S$ term does not enter directly into the running 
of the other soft parameters until three-loop 
order~\cite{Martin:1993zk,Jack:1999zs}, and therefore its effects
on these parameters is expected to be mild.

  There is also a theoretical uncertainty induced by the $S$ term.  
Such a term is effectively equivalent to a Fayet-Iliopoulos~(FI) 
$D$ term for hypercharge~\cite{Jack:1999zs}.  
To see how this comes about, consider the hypercharge 
$D$-term potential including a FI term,
\be
\mathscr{L} = \ldots + \frac{1}{2}D_1^2 + \xi\,D_1 
+ \sqrt{\frac{3}{5}}g_1\,D_1\,\sum_i\bar{\phi}_iY_i\phi^i 
- \sum_i \tilde{m}_i^2|\phi^i|^2,
\ee
where we have also included the soft scalar masses.
Eliminating the $D_1$ through its equation of motion, we find
\beq
\mathscr{L} = \ldots -\frac{1}{2}\xi^2 - \frac{g_1^2}{2}
\left(\sum_i\bar{\phi}_iQ_i\phi^i\right)^2 
- \sum_i\bar{\phi}_i\left(\tilde{m}_i^2+\sqrt{\frac{3}{5}}\,g_1\,
Y_i\,\xi\right)\phi^i.
\eeq
Thus, except for the constant addition to the vacuum energy,
the effect of the FI term is to shift each of the soft squared masses
by an amount 
\beq
\tilde{m}_i^2 \to m_i^2 := \tilde{m}_i^2+\sqrt{\frac{3}{5}}\,g_1\,Y_i\,\xi.
\label{1:mshift}
\eeq  
The low-energy observable quantities are the $\{m_i^2\}$, 
not the $\{\tilde{m}_i^2\}$.  Since we can't extract the shift in 
the vacuum energy, the low-energy effects of the FI term are 
therefore invisible to us, other than the shift in the soft masses.
This shift is exactly the same as the shift due to the $S$ term.
Let us also mention that the $S$ term, as we have defined it in 
Eq.~(\ref{1:sterm}), runs inhomogeneously at two-loops and above,
so the exact correspondence between a hypercharge $D$ term and the $S$ 
term of Eq.~(\ref{1:sterm}) does not hold beyond one-loop order.  

  A simple way to avoid both the large RG uncertainties in the soft
masses and the theoretical ambiguity due to the $S$ term
is apparent from Eq.~(\ref{1:delm2}).  Instead of looking at 
individual soft masses, it is safer to consider the mass differences
\be
Y_j\,m_i^2-Y_i\,m_j^2,
\label{1:diff}
\ee
for any pair of fields.  These differences are not affected by
the mass shifts of Eq.~(\ref{1:delm2}).  They are also independent 
of the value of the FI term.  

  In the rest of this section, we show how the $S$ term 
can complicate the running of the soft masses to high energies
with a particular example.  If one of the scalar soft masses is 
unmeasured, it is essential to use the linear mass combinations 
given in Eq.~(\ref{1:diff}) instead of the individual masses themselves.  
We also discuss how the special RG properties of the $S$ term provide
a useful probe of high scale physics if all the scalar soft masses
are determined experimentally.

\subsection{Example: SPS-5 with an Unmeasured Higgs Soft Mass}

  To illustrate the potential high-scale uncertainties in
the RG-evolved soft parameters due to the $S$ term, 
we study the sample mSUGRA point SPS-5~\cite{Allanach:2002nj}
under the assumption that one of the scalar soft masses
goes unmeasured at the LHC.  If this is the case, the $S$ parameter
is undetermined, and the precise values of the high-scale
soft terms are no longer precisely calculable.  Even if the value of
the $S$ term is bounded by considerations of naturalness,
the uncertainties in the high-scale values of the soft scalar masses 
can be significant. 

  The SPS-5 point is defined by the mSUGRA input values
$m_0 = 150$~GeV, $m_{1/2} = 300$~GeV, $A_0 = -1000$, $\tan\beta = 5$,
and $sgn(\mu) > 0$, at $M_{GUT}$.  The mass spectrum for this 
point has relatively light sleptons around 200~GeV,
and somewhat heavier squarks with masses near 400-600~GeV.
The LSP of the model is a mostly Bino neutralino, with
mass close to 120~GeV.  The perturbation we consider
for this point is a shift in the down-Higgs soft mass, $m_{H_d}^2$.

\begin{figure}[tb]
\begin{center}
\includegraphics[width = 0.65\textwidth]{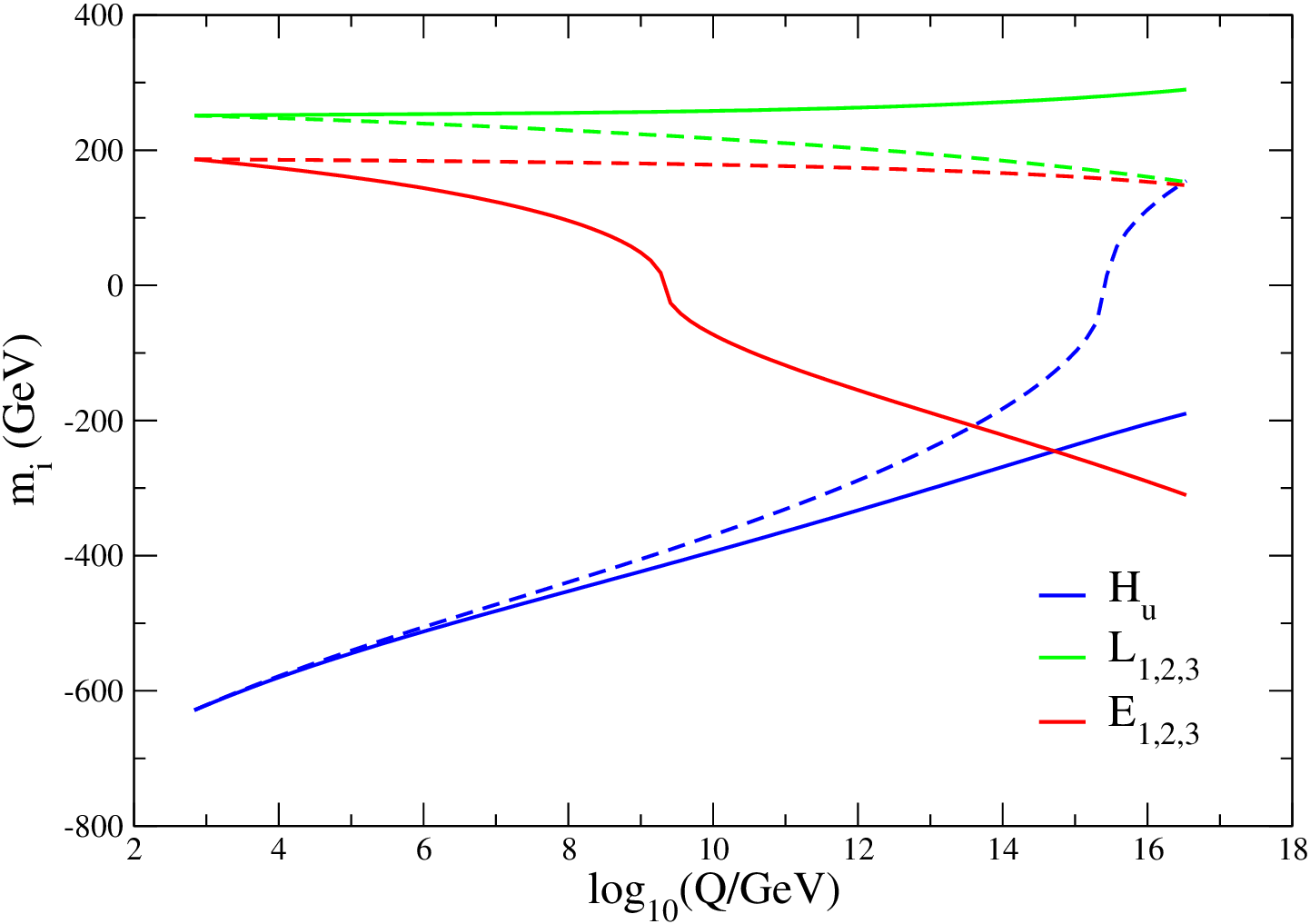}
\end{center}
\caption{\small Deviations in the running of some of the SPS-5 soft
masses due to setting $m_{H_d}^2=(1000~\gev)^2$ at the low scale.
The solid lines show the running of $m_{H_u}^2$, $m_{E}^2$,
and $m_{L}^2$ with this perturbation, while the dashed lines
show the unperturbed running of these soft masses.
The unperturbed low-scale value of the down-Higgs soft mass
is $m_{H_d}^2 \simeq (235\,\gev)^2$.
}
\label{fig:sps5-a}
\end{figure}

  Of the soft supersymmetry breaking parameters in the MSSM,
the soft terms associated with the Higgs sector can be 
particularly difficult to deduce from LHC measurements.  
At tree-level, the independent Lagrangian parameters relevant to this
sector are~\cite{Martin:1997ns}
\beq
v,~~\tan\beta,~~\mu,~~M_A,
\label{hpar}
\eeq
where $v\simeq 174\,\gev$ is the electroweak breaking scale,
$\tan\beta = v_u/v_d$ is the ratio of the $H_u$ and $H_d$ VEVs,
$\mu$ is the supersymmetric $\mu$-term, and $M_A$ is the pseudoscalar
Higgs boson mass.  Other Higgs-sector Lagrangian parameters, 
such as $m_{H_d}^2$ and $m_{H_u}^2$, can be expressed in terms of 
these using the conditions for electroweak symmetry breaking in the MSSM.

\begin{figure}[tb]
\begin{center}
\includegraphics[width = 0.65\textwidth]{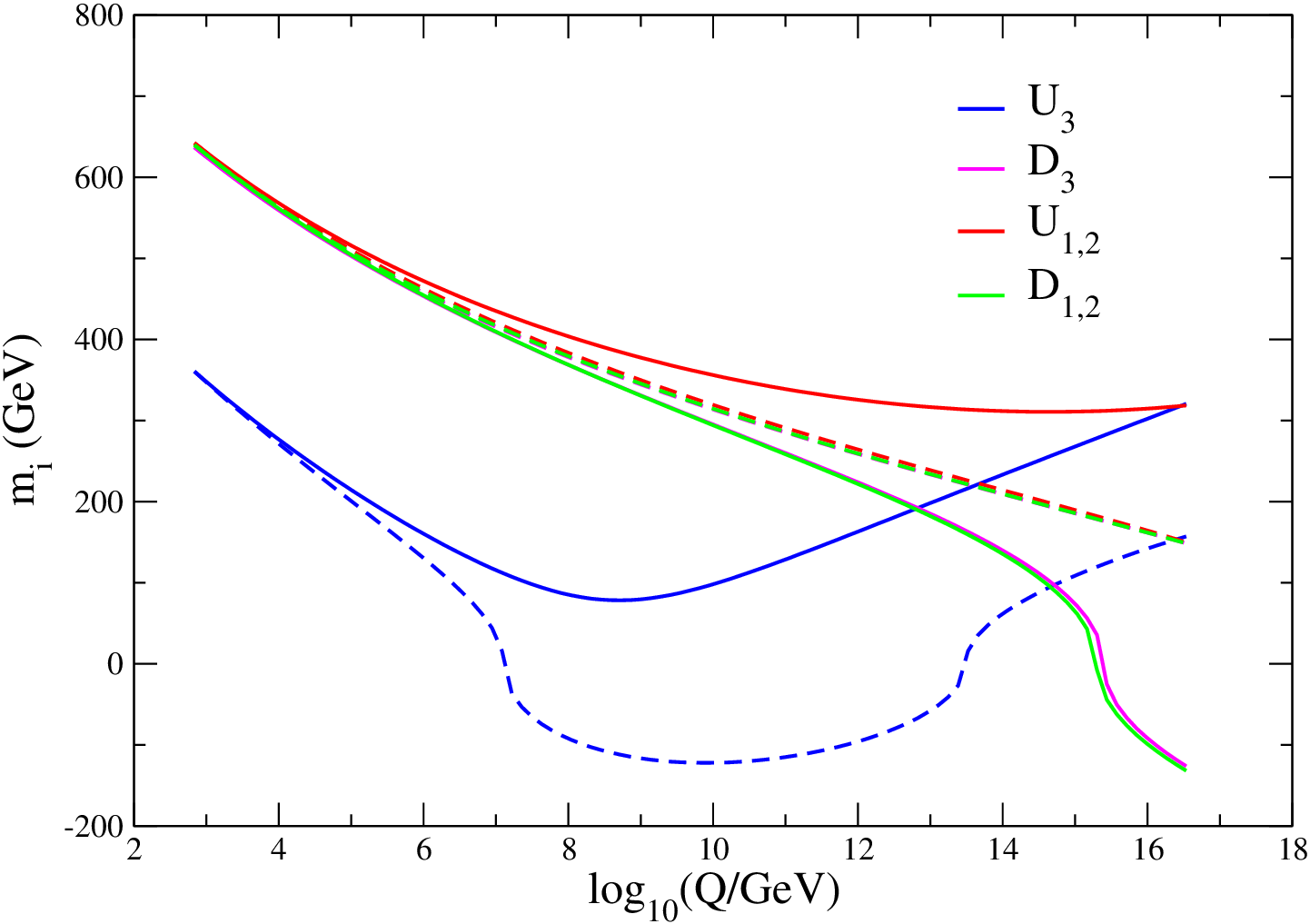}
\end{center}
\caption{\small Deviations in the running of some of the SPS-5 soft
masses due to setting $m_{H_d}^2=(1000~\gev)^2$ at the low scale.
The solid lines show the running of $m_{U_{1,2}}^2$,
$m_{D_{1,2}}^2$, $m_{U_3}^2$, and $m_{D_3}^2$ with this perturbation, 
while the dashed lines show the unperturbed running of these soft masses.}
\label{fig:sps5-b}
\end{figure}

  Among the Higgs sector parameters listed in Eq.~(\ref{hpar}),
only the value of $v$ is known.  
The value of $\mu$ can potentially be studied
independently of the Higgs scalar sector by measuring neutralino 
and chargino masses and couplings~\cite{Zerwas:2002as,Allanach:2002nj,
neutralinomu}, although it is likely to be poorly determined
if only hadron colliders are available.
A number of observables outside the Higgs sector may also be 
sensitive to $\tan\beta$, especially if it is large, $\tan\beta \gtrsim 20$.
For example, the dilepton invariant mass distributions in the 
inclusive $2\ell+jets+\met$ channel can vary significantly depending
on the value of $\tan\beta$, but this dependence is such
that the value of $\tan\beta$ can at best only be confined to 
within a fairly broad ranges~\cite{denegri}.
Determining $M_A$ at the LHC typically requires the discovery
of one of the heavier MSSM Higgs boson states.   
Finding these states can also help to determine 
$\tan\beta$~\cite{Assamagan:2004ji}.
Unfortunately, the LHC reach for the heavier Higgs states is limited,
especially for larger values of $M_A$ and intermediate or smaller values
of $\tan\beta\lesssim 20$~\cite{heavyhiggs}.  If none of the heavier 
Higgs bosons are found at the LHC, it does not appear to be possible 
to determine both of the Higgs soft masses, $m_{H_u}^2$ and $m_{H_d}^2$.

  For the low-scale parameters derived from SPS-5, the pseudoscalar Higgs
mass $M_A$ is about 700~GeV.  With such a large value of
$M_A$, and $\tan\beta = 5$, only the lightest SM-like Higgs boson 
is within the reach of the LHC~\cite{heavyhiggs}.  
Motivated by this observation, we examine the effect of changing 
the low-scale value of $m_{H_d}^2$ on the running of the 
other soft parameters.  The actual low-scale value of $m_{H_d}^2$ 
is about $(235~\gev)^2$.  The perturbation we consider is to set 
this value to $(1000~\gev)^2$, while keeping $\tan\beta$ fixed.
Such a perturbation does not ruin electroweak symmetry breaking,
and tends to push the heavier higgs masses to even larger values.
In the present case, the heavier Higgs masses increase from
about $700~\gev$ to over $1200~\gev$.~\footnote{The values of 
$\mu$ and $B\mu$ also change, although the variation in $\mu$ is very mild:
$\mu\simeq 640\,\gev \to 670\,\gev$.}

  The effects of this perturbation in $m_{H_d}^2$ on some of the soft
scalar masses are shown in Figs.~\ref{fig:sps5-a} and~\ref{fig:sps5-b}.
In these plots, we show $m_i = m_i^2/\sqrt{|m_i^2|}$.
The deviations in the soft masses are substantial, and the 
$S$ parameter is the source of this uncertainty.
In addition to the $S$ term, varying $m_{H_d}^2$ can also modify 
the running of the Higgs mass parameters and the down-type squarks
and sleptons through the Yukawa-dependent terms $X_i$ in the 
RG equations, Eq.~(\ref{1:msoft}).  
In the present case these Yukawa-dependent effects are very mild
since for $\tan\beta = 5$, the $b$ and $\tau$ Yukawas are still quite small.  
This can be seen by noting the small difference between the 
perturbed running of $m_{U_{1,2}}^2$ and $m_{U_3}^2$ in Fig.~\ref{fig:sps5-b}.
For larger values of $\tan\beta$, the down-type Yukawa couplings can be 
enhanced and this non-$S$ effect from $m_{H_d}^2$ can be significant.  
However, we also note that as these Yukawa couplings grow larger, 
it becomes much easier for the LHC to detect one or more of the heavier 
Higgs states.  To the extent that the Yukawa-dependent shifts 
can be neglected, the linear mass combinations of Eq.~(\ref{1:diff}) 
remove most of the uncertainty due to an unknown $m_{H_d}^2$ 
in the running of the soft masses that are measured.  
The effect of not knowing $m_{H_d}^2$ has only a very small 
effect on the running of the gaugino masses and the trilinear terms.

  In this example we have assumed that $m_{H_d}^2$ is the only
unmeasured soft scalar mass.  Several of the other soft scalar
masses may be difficult to reconstruct from LHC data as well.
For example, within many SUSY scenarios the third generation squarks 
and some of the heavier sleptons have very small LHC production rates. 
If there are other unmeasured soft scalar masses besides $m_{H_d}^2$, 
the uncertainties due to the $S$ term in the extrapolation of the measured 
scalar soft masses will be even greater than what we have presented here.
The mass combinations of Eq.~(\ref{1:diff}) will be necessary
to study the high scale supersymmetry breaking spectrum in
this case.

\subsection{Origins and Uses of the $S$ Term}

  While the $S$ term can complicate the extrapolation 
of the soft scalar masses if one of them goes unmeasured, 
the simple scale dependence of this term also makes it a useful
probe of the high scale theory if all the masses are determined.
The essential feature is the homogeneous RG evolution of the $S$ term, 
given in Eq.~(\ref{1:dels}), which is related to the non-renormalization 
of FI terms in the absence of supersymmetry breaking.

  A non-vanishing $S$ term can arise from a genuine FI term 
present in the high scale theory, or from non-universal scalar 
soft masses at the high input scale.  The size of a fundamental
hypercharge FI term, $\xi$, is na\"ively on the order of 
the large input scale.  Such a large value would either destabilize 
the gauge hierarchy or lead to $U(1)_Y$ breaking at high energies.  
However, the non-renormalization of FI terms implies that 
it is technically natural for $\xi$ to take on much smaller values.  
In this regard, a small value for $\xi$ is analogous to the 
$\mu$ problem in the MSSM.  Adding such a FI term to mSUGRA
provides a simple one-parameter extension of this model, and can
have interesting effects~\cite{deGouvea:1998yp,Falk:1999py}.
Note, however, that in a GUT where $U(1)_Y$ is embedded 
into a simple group, a fundamental hypercharge FI term 
in the full theory is forbidden by gauge invariance.

  It is perhaps more natural to have the $S$ term emerge from
non-universal scalar soft masses~\cite{dterm,dgut}.  
This is true even in a GUT where $U(1)_Y$ is embedded into a simple group.  
Within such GUTs, the contribution to $S$ from complete 
GUT multiplets vanishes.
However, non-zero contributions to $S$ can arise from multiplets
that are split in the process of GUT breaking.  For example, in $SU(5)$ 
with $H_u$ and $H_d$ states embedded in $\bf{5}$ and $\overline{\bf{5}}$ 
multiplets, a non-zero low-energy value of $S$ can be generated when 
the heavy triplet states decouple provided the soft masses 
of the respective multiplets are unequal.
Whether it is zero or not, the low-scale value of the $S$ term
provides a useful constraint on the details of 
a GUT interpretation of the theory.

  So far we have only considered the $S$ term corresponding to
$U(1)_Y$.  If there are other gauged $U(1)$ symmetries, there will
be additional $S$ term-like factors for these too.  In fact, 
the homogeneity of the $S$ term evolution also has a useful 
implication for any non-anomalous \emph{global} $U(1)$ symmetry 
in the theory.  The only candidate in the MSSM is $U(1)_{B-L}$,
up to linear combinations with $U(1)_Y$~\cite{u1x}.  Let us define
$S_{B-L}$ by the combination
\bea
S_{B-L} &=& Tr(Q_{B-L}m^2)\label{1:sbl}\\
&=& tr(2m_Q^2 - m_U^2-m_D^2 - 2m_L^2+m_E^2),\nnmb
\eea
where the second trace runs only over flavors.\footnote{
Up to flavor mixing, we can also define an independent 
$S_{B-L}$ for each generation.}
We can think of $S_{B-L}$ as the effective $S$ term for a
gauged $U(1)_{B-L}$ in the limit of vanishing coupling.
At one loop order, the RG running of $S_{B-L}$ is given by
\bea
(16\pi^2)\frac{dS_{B-L}}{dt} &=& \frac{3}{5}Tr(Q_{B-L}Y)g_1^2\,S\label{1:bls}\\
&=& n_g\frac{16}{5}\,g_1^2\,S,\nnmb
\eea
where $n_g=3$ is the number of generations.
If $S$ is measured and vanishes, $S_{B-L}$ provides a second useful
combination of masses that is invariant under RG evolution,\footnote{
This non-evolution of mass combinations corresponding to 
non-anomalous global symmetries persists at strong coupling.
In models of conformal sequestering, 
this can be problematic~\cite{Luty:2001jh}.}
and yields an additional constraint on possible GUT embeddings
of the theory.


\section{New Physics: Complete GUT Multiplets\label{gutmult}}

  As a second line of investigation, we consider the effects of
some possible types of new intermediate scale physics 
on the running of the MSSM soft terms.  If this new physics 
is associated with supersymmetry breaking as in 
gauge mediation~\cite{Dine:1993yw},
then it is of particular interest in its own right.
Indeed, in this case the low-energy spectrum of soft terms may 
point towards the identity of the new physics after RG evolution.
On the other hand, there are many kinds of possible new intermediate 
scale physics that are not directly related to supersymmetry breaking.
The existence of this type of new phenomena can make it much more
difficult to deduce the details of supersymmetry breaking from
the low-energy soft terms.  

  A useful constraint on new physics is gauge coupling unification.
To preserve unification, the SM gauge charges of the new physics
should typically be such that all three MSSM gauge beta functions are
modified in the same way.\footnote{For an interesting exception,
see Ref.~\cite{Martin:1995wb}.}  This is automatic if 
the new matter fills out complete multiplets of a simple GUT group 
into which the MSSM can be embedded.  Such multiplets can emerge 
as remnants of GUT symmetry breaking. 
 
  As an example of this sort of new physics, we consider vector-like 
pairs of complete $SU(5)$ multiplets.  For such multiplets, it is possible
to write a down a supersymmetric mass term of the form
\be
\mathcal{W} \supset \tilde{\mu}\,\bar{X}\,X,
\ee
where $X$ and $\bar{X}$ denote the chiral superfields of the 
exotic multiplets.  We also assume that the exotic multiplets
have no significant superpotential (Yukawa) couplings with the 
MSSM fields.  Under these assumptions, the exotic $SU(5)$ multiplets can
develop large masses independently of the details of the MSSM.
They will interact with the MSSM fields only through their 
gauge interactions.

  If the supersymmetric mass $\tilde{\mu}$ is much larger than the
electroweak scale, it will be very difficult to deduce the
presence of the additional GUT multiplets from low-energy data alone.
Moreover, an extrapolation of the measured soft masses using the 
RG equations appropriate for the MSSM will lead to incorrect values
of the high-scale parameters.  In this section, we characterize
the sizes and patterns of the deviations in the high scale soft
spectrum induced by additional vector-like GUT multiplets.
We also show that even though the new matter interferes with the 
running of the MSSM soft parameters, it is often still possible to 
obtain useful information about the input spectrum, such as the 
relative sizes of the gaugino masses and whether there 
is inter-generational splitting between the soft scalar masses.

\subsection{Shifted Gauge Running}

  The main effect of the exotic GUT multiplets is to shift the running
of the $SU(3)_c$, $SU(2)_L$, and $U(1)_Y$ gauge couplings.  Recall that
in the MSSM, the one-loop running of these couplings is determined by
\beq
\frac{dg_a^{-2}}{dt} = \frac{b_a}{8\pi^2},
\label{rungauge}
\eeq
with $(b_1,~b_2,~b_3) = (-33/5,-1,3)$.  
The presence of a massive GUT multiplet shifts each of the $b_a$ up 
by an equal amount above the heavy threshold at 
$t = t_I = \ln(\tilde{\mu}/M_Z)$,
\beq
\Delta b = - \nfive - 3\,N_{{\bf{10}\oplus\bar{\bf 10}}} + \ldots
\eeq
where $\nfive$ is the number of additional $\bf{5}\!\oplus\!\overline{\bf{5}}$ 
representations and $N_{{\bf{10}\oplus\bar{\bf 10}}}$ is the number of 
$\bf{10}\oplus\overline{\bf{10}}$'s.
The modified one-loop solution to the RG equations is therefore,
\beq
g_a^{-2}(t) = \left\{
\begin{array}{lcc}
g_a^{-2}(t_0) +\frac{b_a}{8\pi^2}(t-t_0)&~~~~~&t<t_I,\\
g_a^{-2}(t_0) +\frac{b_a}{8\pi^2}(t-t_0)+\frac{\db}{8\pi^2}(t-t_I)
&~~~~~&t>t_I.
\end{array}\right.\label{grun}
\eeq
It follows that the unification scale is not changed, but the value 
of the unified gauge coupling is increased.  Note that the 
number of new multiplets is bounded from above for a given value
of $\tilde{\mu}$ if gauge unification is to be perturbative.\footnote{
Note that the parts of $\tilde{\mu}$ corresponding to the doublet
and triplet components of the $\bf{5}$ will run differently between 
$M_{GUT}$ and the intermediate scale.  This will induce additional threshold
corrections that we do not include in our one-loop analysis.}

  The change in the gauge running shifts the running of all the soft
parameters, but the greatest effect is seen in the gaugino masses.
At one-loop, these evolve according to
\beq
\frac{dM_a}{dt} = -\frac{b_a}{8\pi^2}g_a^2\,M_a.
\label{rungino}
\eeq
It follows that $M_a/g_a^2$ is RG invariant above and below the heavy 
threshold.  If the threshold is also supersymmetric, $M_a$ will
be continuous across it at tree-level.  Since $g_a$ is also
continuous across the threshold at tree-level, the addition of 
heavy vector-like matter does not modify
the one-loop scale invariance of the ratio $M_a/g_a^2$.
This holds true whether or not the new matter preserves gauge
unification, but it is most useful when unification holds.
When it does, the measurement of the low-energy gaugino masses
immediately translates into a knowledge of their ratio at 
$M_{GUT}$~\cite{Kane:2002ap}.  

\begin{figure}[tbh]
\vspace{1cm}
\begin{center}
\includegraphics[width = 0.65\textwidth]{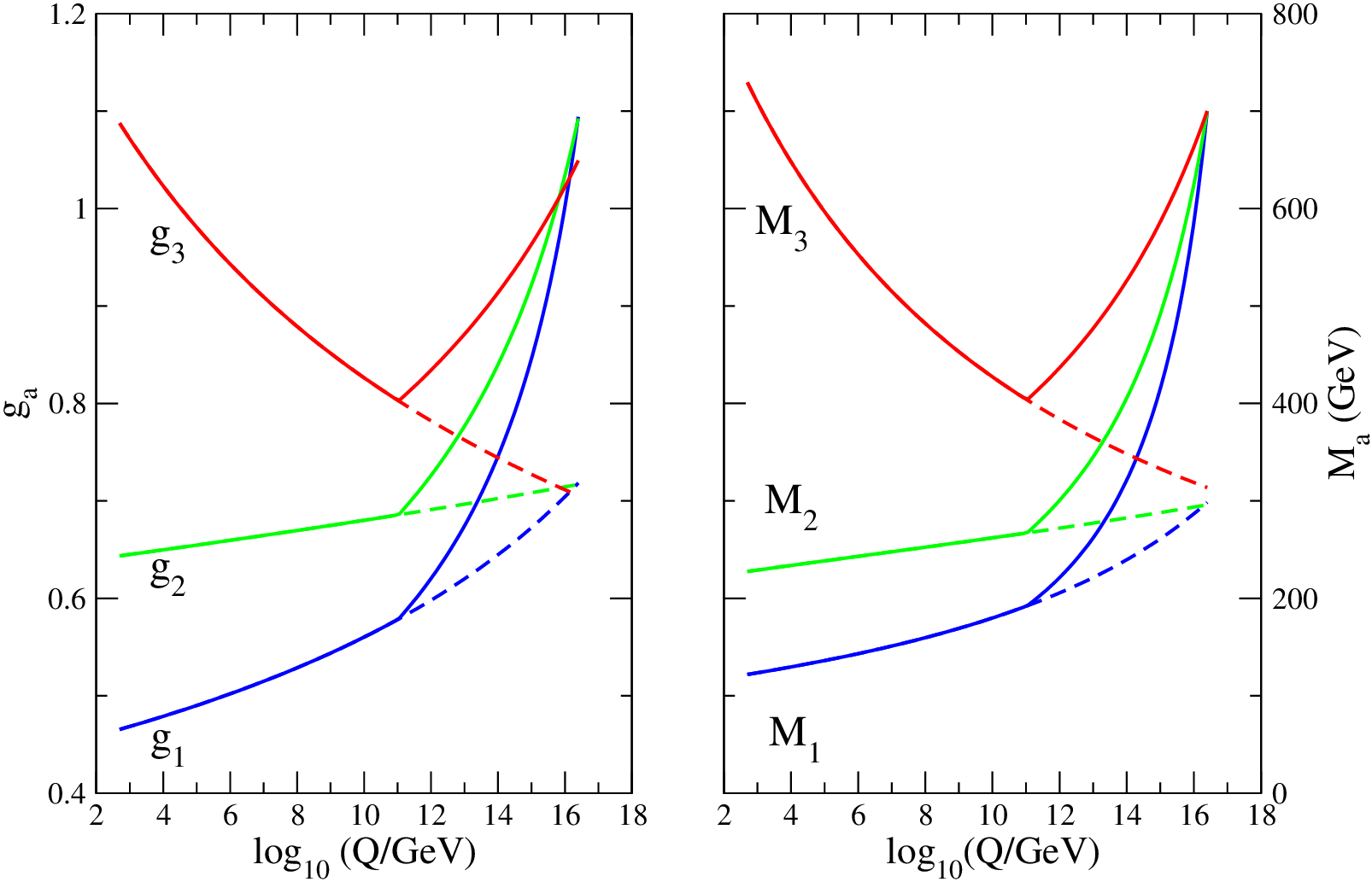}
\end{center}
\caption{\small The shift in the running of the gauge couplings~(left) and the 
gaugino masses~(right) due to 7 sets of ${\bf 5}\oplus\bar{\bf 5}$'s
with mass $\tilde{\mu} = 10^{11}$~GeV.  The universal gaugino mass
is taken to be $700\,\gev$.} 
\label{fig:gshift}
\end{figure}

  From Eq.~(\ref{grun}) and the one-loop scale invariance
of $M_a/g_a^2$, the shift in the gaugino masses due to the 
additional matter is
\beq
M_a(t) = \bar{M}_a(t)\,\left[1+\frac{\Delta b\,\bar{g}_a^2}{8\pi^2}
(t-t_I)\right]^{-1},
\eeq
where $\bar{g}_a$ and $\bar{M}_a$ denote the values these parameters
would have for $\Delta b = 0$ (\emph{i.e.} the values obtained using
the MSSM RG equations).  For $t=t_{GUT}$, the shift coefficient 
is identical for $a=1,2,3$ provided gauge unification occurs.
The shift in the running of the gauge couplings and the gaugino 
masses due to seven sets of ${\bf 5}\oplus\bar{\bf 5}$'s 
at $10^{11}$~GeV is illustrated in Fig.~\ref{fig:gshift}.
An unperturbed universal gaugino mass of $M_{1/2} = 700\,\gev$ 
is assumed.  Both the values of the unified gauge coupling
and the universal gaugino mass at $M_{GUT}$ are increased
by the additional multiplets.

  The running of the soft masses also depends on the running of the 
gauge couplings and gaugino masses, and is modified by the appearance 
of new matter.  At one-loop order, in the limit that we can neglect 
the Yukawa couplings, it is not hard to find the shifts in the soft masses.  
We can divide these shifts into two contributions,
\beq
m_i^2(t) = \bar{m}_i^2 + \Delta m^2_{i_{\lambda}} + \Delta m^2_{i_{S}},
\eeq
where $\bar{m}_i^2$ is the value of the soft mass obtained by running
the measured value up in the absence of the new matter, 
$\Delta m^2_{i_{\lambda}}$ is the shift due to the modified gaugino masses,
and $\Delta m^2_{i_{S}}$ is due to the change in the running of the $S$ term.

  The first shift, $\Delta m^2_{i_{\lambda}}$, 
can be obtained straightforwardly
from Eqs.~(\ref{1:msoft},\ref{rungauge},\ref{rungino}), and is given by
\beq
\Delta m_{i_{\lambda}}^2 = \sum_a2\,C^a_i\left|\frac{M_a}{g_a^2}\right|^2
\Delta I_a
\label{delml}
\eeq
where $C^a_i$ the Casimir invariant of the $a$-th gauge group and
\beq
\Delta I_a = \frac{1}{(b_a+\db)}\,(g_a^4-g_{a_I}^4) 
- \frac{1}{b_a}(\bar{g}_a^4-g_{a_I}^4),
\eeq
with $g_a$ the actual gauge coupling at scale $t>t_I$ (including
the extra matter), $g_{a_I}$ the gauge coupling at the heavy threshold $t_I$,
and $\bar{g}_a$ the gauge coupling at scale $t>t_I$ in the absence
of new matter.

  New chiral matter can also modify the running of the soft scalar masses
through the $S$ parameter.  The new matter shifts the running of $S$
by changing the running of $g_1$, but it can also contribute to the 
$S$ term directly at the heavy threshold.      
Combining these effects, the value of $S$ above threshold is
\bea
S(t) &=& \left(\frac{{g}_1}{g_{1_0}}\right)^2\,S(t_0) 
+ \left(\frac{g_1}{g_{1_I}}\right)^2\,\Delta\,S\\
&=& \bar{S}(t) + \left(\frac{{g}_1^2-\bar{g}_1^2}{g_{1_0}^2}\right)\,S(t_0) 
+ \left(\frac{g_1}{g_{1_I}}\right)^2\,\Delta\,S,\nnmb
\eea
where $\Delta S$ is the shift in the value of $S$ at the threshold,
$g_{1_0}$ is the low-scale value of the gauge coupling,
and $\bar{S}(t)$ is the value the $S$ term would have in the absence
of the new matter.
Inserting this result into Eq.~(\ref{1:msoft}), the effect of the new
matter on the running of the soft scalar masses through the $S$ term is
\bea
\frac{5}{3Y_i}\Delta m^2_{i_S} &=& 
\frac{1}{b_1}\left[
\left(\frac{\bar{g}_{1}}{g_{1_0}}\right)^2 
- \left(\frac{g_{1_I}}{g_{1_0}}\right)^2\right]\,S_0
- \frac{1}{(b_1+\db)}\left[
\left(\frac{g_{1}}{g_{1_0}}\right)^2 
- \left(\frac{g_{1_I}}{g_{1_0}}\right)^2\right]\,S_0\nnmb\\
&&\label{delms}\\
&&~~~- \frac{1}{(b_1+\db)}\left[
\left(\frac{g_{1}}{g_{1_I}}\right)^2 -1\right]\,\Delta S.\nnmb
\eea  
As before, the effects of the $S$ term on the soft masses are
universal up to the hypercharge prefactor.  Thus, they still
cancel out of the linear combinations given in Eq.~(\ref{1:diff}).

\subsection{Yukawa Effects and Useful Combinations}

  We have so far neglected the effects of the MSSM Yukawa
couplings on the modified running of the soft masses.  As a result,
the shifts in the running of the soft scalar masses written above
are family universal.  There are also non-universal shifts in
the soft scalar masses.  These arise from the Yukawa-dependent
terms in the soft scalar mass beta functions, which themselves
depend on the Higgs and third-generation soft scalar masses.
As a result, the low-energy spectrum obtained
from a theory with universal scalar masses at the high scale
and additional GUT multiplets can appear to have non-universal
soft masses at the high scale if the extra GUT multiplets 
are not included in the RG evolution.
These non-universal shifts are usually a subleading effect, 
but as we illustrate below they can be significant when 
the supersymmetric mass of the new GUT multiplets is within 
a few orders of magnitude of the $\tev$ scale.

  Non-universal mass shifts obscure the relationship
between the different MSSM generations and the source 
of supersymmetry breaking.  This relationship is closely linked to 
the SUSY flavor problem~\cite{superckm}, and possibly also to
the origin of the Yukawa hierarchy.  For example, third generation 
soft masses that are significantly different from the first and 
second generation values is one of the predictions of the model 
of Ref.~\cite{Nelson:2000sn}, in which strongly-coupled 
conformal dynamics generates the Yukawa hierarchy and suppresses 
flavor-mixing soft terms.  The relative sizes of the high scale
soft masses for different families is therefore of great 
theoretical interest.
  
  Even when there are non-universal shifts from new physics, 
it is sometimes still possible to obtain useful information about 
the flavor structure of the soft scalar masses at the high scale.
To a very good approximation, the flavor non-universal 
contributions to the RG evolution of the soft masses are proportional to
the third generation Yukawa couplings or the trilinear couplings.
There is also good motivation (and it is technically allowed) 
to keep only the trilinear couplings for the third generation.  
In this approximation, the Yukawa couplings and the $A$ terms only appear 
in the one-loop RG equations for the soft masses through 
three independent linear combinations.  Of the seven soft masses 
whose running depends on these combinations, 
we can therefore extract four mass combinations
whose evolution is independent of Yukawa effects at one-loop 
order~\cite{iblop}.\footnote{
We also assume implicitly that the soft masses are close to
diagonal in the super CKM basis, as they are quite 
constrained to be~\cite{superckm}.}
They are:
\bea
m^2_{A_3} &=& 2\,m_{L_3}^2-m_{E_3}^2\label{massflav1}\\
m^2_{B_3} &=& 2\,m_{Q_3}^2 - m_{U_3}^2 - m_{D_3}^2\nnmb\\
m^2_{X_3} &=& {2}\,m_{H_u}^2 - 3\,m_{U_3}^2\nnmb\\
m^2_{Y_3} &=& {3}\,m_{D_3}^2 + 2\,m_{L_3}^2 - 2\,m_{H_d}^2\nnmb
\eea
The cancellation in the first two terms occurs because the linear 
combinations of masses correspond to $L$ and $B$
global symmetries.  They run only because these
would-be symmetries are anomalous under $SU(2)_L$ and $U(1)_Y$.
The other mass combinations can also be related to anomalous global
symmetries of the MSSM.

  These mass combinations have the same
one-loop RG running as certain combinations of masses involving 
only the first and second generations.  For example, 
the $m_{B_3}^2$ combination runs in exactly the same way at one-loop as 
\beq
m_{B_{i}}^2 = 2\,m^2_{Q_{i}} - m^2_{U_{i}}-m^2_{D_{i}},
\label{massflav2}
\eeq 
for $i=1,2$.
If these linear combinations are unequal at the low scale,
the corresponding soft masses will be non-universal at the high scale.
This holds in the MSSM, as well as in the presence of any new physics
that is flavor universal and respects baryon number.  
On the other hand, $m^2_{B_3} = m^2_{B_1}$ does not imply 
family-universal high scale masses.  For example, 
within a $SO(10)$ GUT, a splitting between the soft masses 
of the $\bf{16}$'s containing the first, second, and third generations 
will not lead to a difference between the mass combinations 
in Eq.~(\ref{massflav1}) at the low-scale.
A similar conclusion holds for the mass combinations $m^2_{A_i}$.

  In the case of $m^2_{X_3}$ and $m^2_{Y_3}$, it is less obvious
what to compare them to.  The trick here is to notice that
in the absence of Yukawa couplings, $m^2_{H_d}$ runs in the same way
as $m_{L_1}^2$ since they share the same gauge quantum numbers.  
If the $S$ term vanishes as well, $m^2_{H_u}$ also has the same RG evolution
as $m^2_{L_1}$.  This motivates us to define
\bea
m_{X_i}^2 &=& {2}\,m_{L_i}^2 - 3\,m_{U_i}^2,\label{massflav3}\\
m_{Y_i}^2 &=& {3}\,m_{D_i}^2,\nnmb
\eea
for $i=1,2$.  These mass combinations can be compared with $m_{X_3}^2$
and $m_{Y_3}^2$ in much the same way as for $m_{B_i}^2$ and $m_{A_i}^2$
(although comparing the $m_{X_i}^2$'s is only useful for $S=0$).
They also correspond to anomalous global symmetries in the limit
that the first and second generation Yukawa couplings vanish.

  The mass combinations listed above can be useful if there
is heavy new physics that hides the relationships between 
the high scale masses.  For instance, suppose the high scale soft
spectrum obtained using the MSSM RG equations applied to 
the  measured soft scalar masses shows a large 
splitting between $m_{Q_3}^2$ and $m_{Q_1}^2$.
If the corresponding splitting between $m_{B_3}^2$ 
and $m_{B_1}^2$ (at any scale) is very much smaller,
this feature suggests that there is new physics that should have
been included in the RG running, or that there exists 
a special relationship between $m_{Q_i}^2$, $m_{U_i}^2$, and $m_{D_i}^2$ 
at the high scale.  A similar conclusion can be made for the 
other mass combinations.

\subsection{Some Numerical Results}

  In our numerical analysis, we follow a similar procedure
to the one used in the previous section.  The MSSM running
is performed at two-loop order, and we interface with
Suspect~2.3.4~\cite{Djouadi:2002ze} to compute the low-scale 
threshold corrections.
New physics, in the form of vector-like GUT multiplets at an intermediate
scale is included only at the one-loop level.  Unlike the previous section, 
we define our high-energy spectrum using a simple mSUGRA
model in the $\Delta b\neq 0$ theory, and include the new physics effects
in generating the low-energy spectrum.  We then evolve this spectrum back up
to the unification scale using the MSSM RG evolution, with $\Delta b =0$.
Our goal is to emulate evolving the MSSM soft parameters
in the presence of unmeasured and unknown high-scale new physics.  

\begin{figure*}[tbh]
\begin{center}
\vspace{1cm}
        \includegraphics[width = 0.65\textwidth]{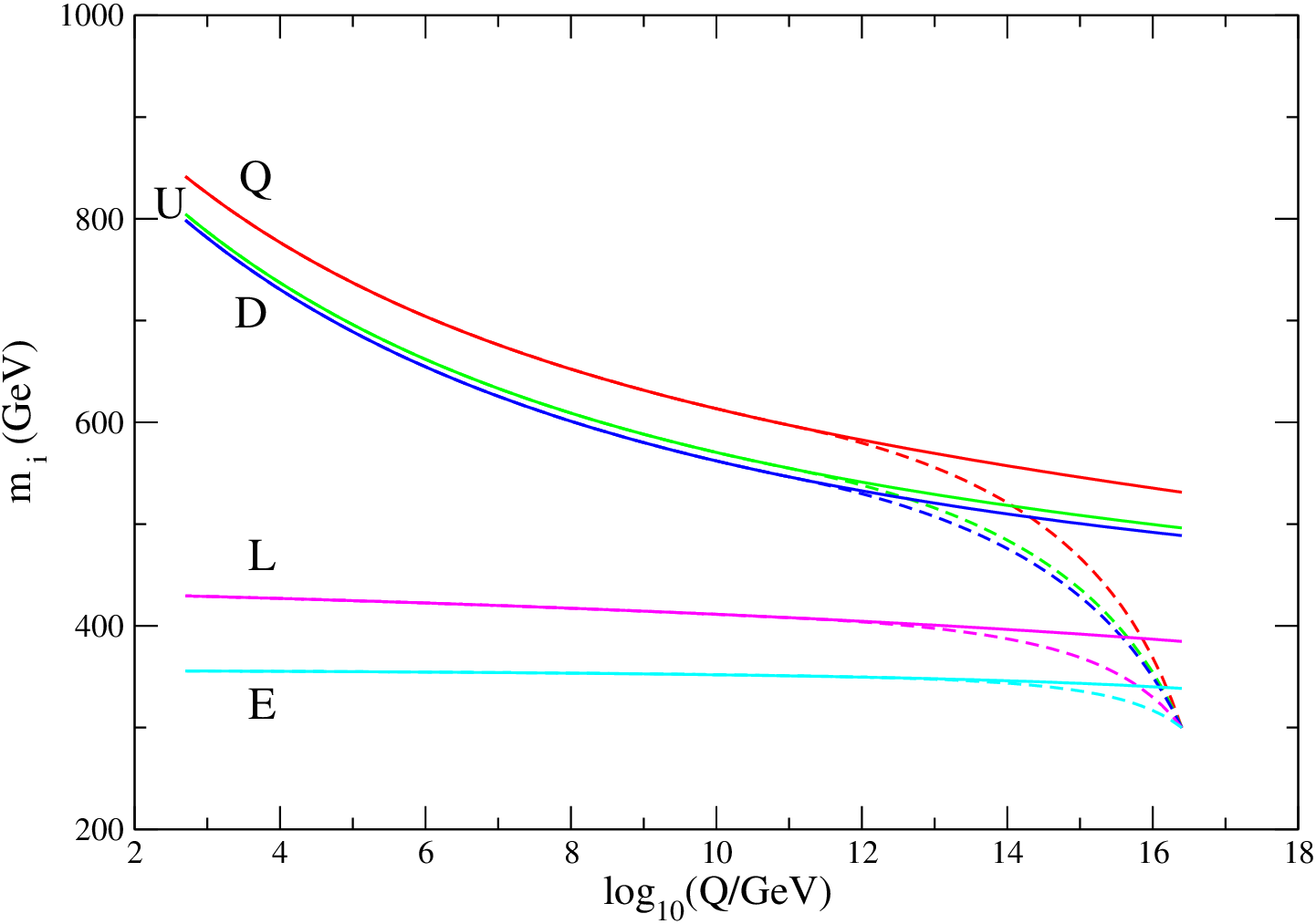}
\caption{\small Running of the first generation soft scalar masses 
with $\nfive = 7$ and $\tilde{\mu} = 10^{11}\,\gev$ for the 
mSUGRA input parameters $m_0 = 300\,\gev$, $m_{1/2} = 700\,\gev$, 
$\tan\beta = 10$, and $A_0 = 0$.  The dashed lines show the actual 
running of these parameters, while the solid lines show the
running from low to high using the RG equations of the MSSM,
ignoring the additional heavy multiplets.} 
\label{fig:n5=7}
\end{center}
\end{figure*}

  The running of the soft masses with $\nfive = 7$ sets of $\five$
multiplets with an intermediate scale mass of $\tilde{\mu} = 10^{11}\,\gev$
is shown in Fig.~\ref{fig:n5=7}
for the mSUGRA parameters $m_0 = 300\,\gev$, $m_{1/2} = 700\,\gev$,
$\tan\beta = 10$, and $A_0 =0$.
These parameters are used to find the low energy spectrum, which is
then RG evolved back up to the high scale with $\Delta b = 0$.
From this figure, we see that a na\"ive MSSM extrapolation 
of the soft parameters (\emph{i.e.} with $\Delta b=0$) yields 
predictions for the high-scale soft scalar masses that are significantly
larger than the actual values.  In the same way, the MSSM predicted values
of the high-scale gaugino masses are smaller than the correct values, 
as can be seen in Fig.~\ref{fig:gshift}.  Note that for $\tilde{\mu} =
10^{11}\,\gev$, $\nfive = 7$ is about as large as possible while
still keeping the gauge couplings perturbatively small all the way
up to $M_{GUT}$.  

\begin{figure*}[tbh]
\begin{center}
\vspace{1cm}
        \includegraphics[width = 0.65\textwidth]{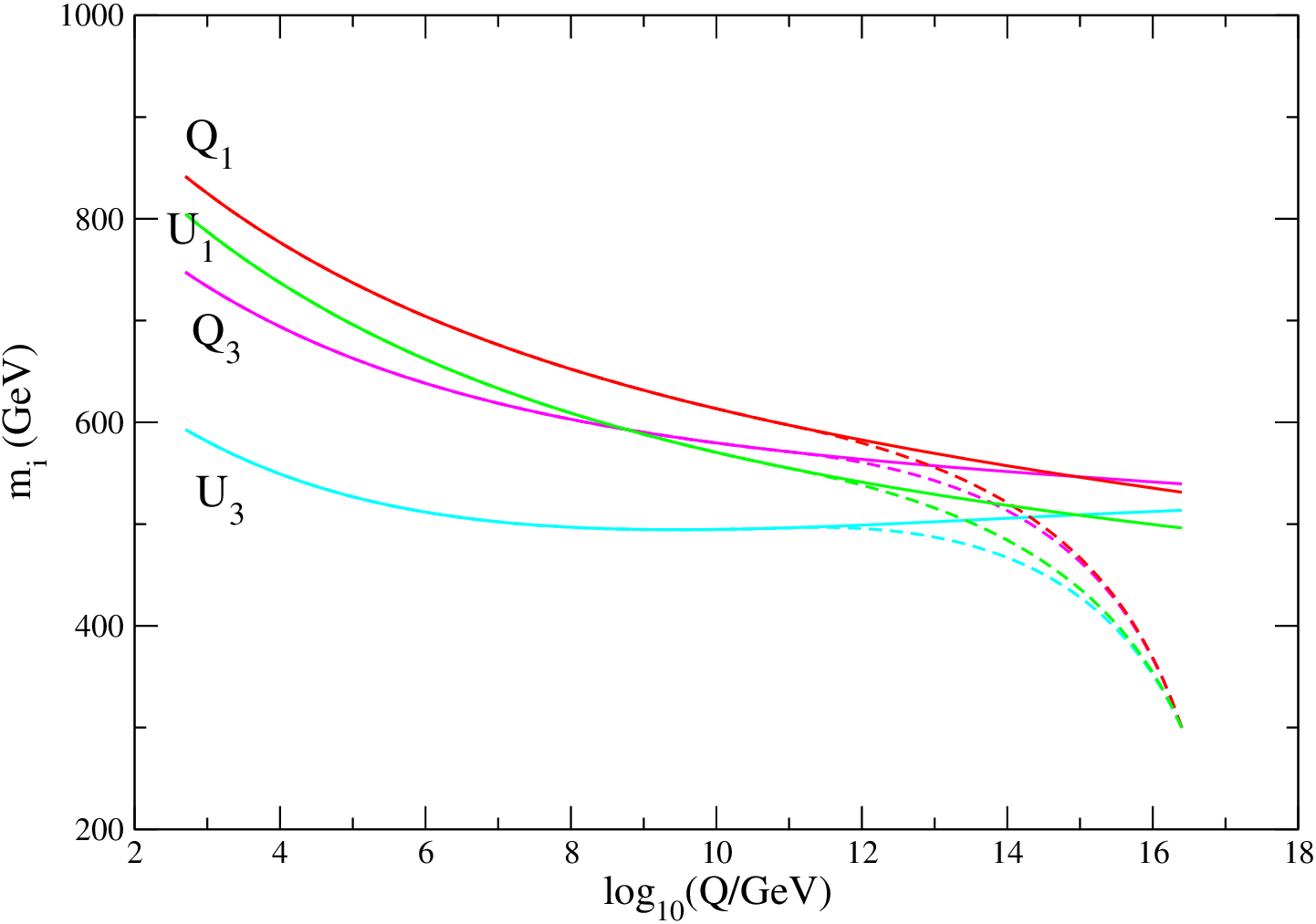}
\caption{\small Running of the soft scalar masses of $Q_{1,3}$ and $U_{1,3}$ 
with $\nfive = 7$ and $\tilde{\mu} = 10^{11}\,\gev$ for the 
mSUGRA input parameters $m_0 = 300\,\gev$, $m_{1/2} = 700\,\gev$, 
$\tan\beta = 10$, and $A_0 = 0$.  The dashed lines show the actual 
running of these parameters, while the solid lines show the
running from low to high using the RG equations of the MSSM,
ignoring the additional heavy multiplets.
} 
\label{fig:univ1}
\end{center}
\end{figure*}

  Besides confusing the relationship between the high scale gaugino
and scalar soft masses, heavy GUT multiplets can also obscure the
comparison of the high scale scalar masses from different generations.
As discussed above, this arises from the backreaction in the 
Yukawa-dependent terms in the RG equations for the third generation
soft scalar masses.  Numerically, we find that the splitting is
quite small compared to the absolute scale of the masses for
$\tilde{\mu} \gtrsim 10^{11}\,\gev$.  This is illustrated
in Fig.~\ref{fig:univ1}.  
We also find that an approximate preservation of universality persists 
for other values of $m_0$, $A_0$, and $\tan\beta$ as well.    
The reason for this appears to be that for $\tilde{\mu} 
\gtrsim 10^{11}~\gev$, the Yukawa couplings are smaller than the
gauge couplings by the time the new physics becomes relevant.

\begin{figure*}[tbh]
\begin{center}
\vspace{1cm}
        \includegraphics[width = 0.65\textwidth]{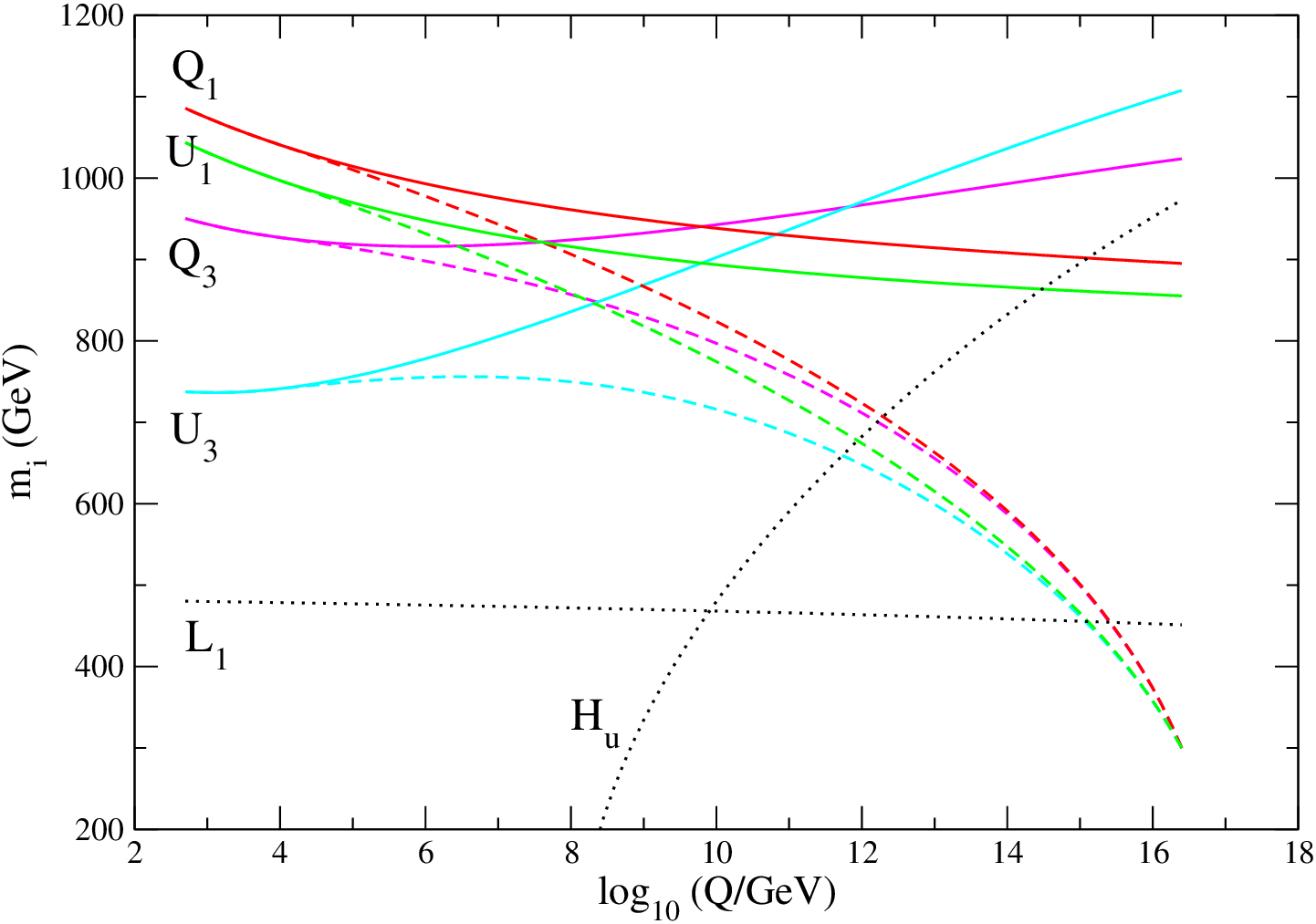}
\caption{\small Running of the soft scalar masses of $Q_{1,3}$ and $U_{1,3}$ 
with $\nfive = 3$ and $\tilde{\mu} = 10^{4}\,\gev$ for the 
mSUGRA input parameters $m_0 = 300\,\gev$, $m_{1/2} = 700\,\gev$, 
$\tan\beta = 10$, and $A_0 = 0$.  The dashed lines show the actual 
running of these parameters, while the solid lines show the
running from low to high using the RG equations of the MSSM,
ignoring the additional heavy multiplets.
} 
\label{fig:univ2}
\end{center}
\end{figure*}  

  A much greater splitting between the high scale values of 
$m_{Q_1}^2$ and $m_{Q_3}^2$, and $m_{U_1}^2$ and $m_{U_3}^2$, 
is obtained for lower values of $\tilde{\mu}$.  
This effect is shown in Fig.~\ref{fig:univ2} for 
$\nfive = 3$ sets of $\five$ multiplets with an intermediate scale mass 
of $\tilde{\mu} = 10^{4}\,\gev$, and the mSUGRA parameters 
$m_0 = 300\,\gev$, $m_{1/2} = 700\,\gev$, $\tan\beta = 10$, and $A_0 =0$.
The value of the gauge couplings at unification here is very
similar to the $\tilde{\mu} = 10^{11}\,\gev$ and $\nfive = 7$ case.
As might be expected, the Yukawa-dependent terms in the soft scalar 
mass RG running become important at lower scales where the top Yukawa 
approaches unity.  

  It is also interesting to note that in both
Figs.~\ref{fig:univ1} and \ref{fig:univ2}, the soft masses appear to
take on family universal values, $m_{Q_1}^2 = m_{Q_3}^2$ 
and $m_{U_1}^2 = m_{U_3}^2$, at the same scale, near $10^{15}\,\gev$ in 
Fig.~\ref{fig:univ1}, and close to $10^{10}\,\gev$ in Fig.~\ref{fig:univ2}.
It is not hard to show, using the mass combinations 
in Eqs.~(\ref{massflav1},\ref{massflav2},\ref{massflav3}),
that this feature holds exactly at one-loop order provided $S=0$,
the high scale masses are family-universal, 
and the only relevant Yukawa coupling is that of the top quark.
In this approximation, all the family-dependent mass splittings are
proportional to $(m_{H_u}^2\!-\!m_{L_1}^2)$, and hence vanish
when $m_{H_u}^2 = m_{L_1}^2$.  This relationship can be seen
to hold approximately in Fig.~\ref{fig:univ2}, which also includes
two-loop and bottom Yukawa effects.

  \begin{figure*}[tbh]
\begin{center}
\vspace{1cm}
        \includegraphics[width = 0.65\textwidth]{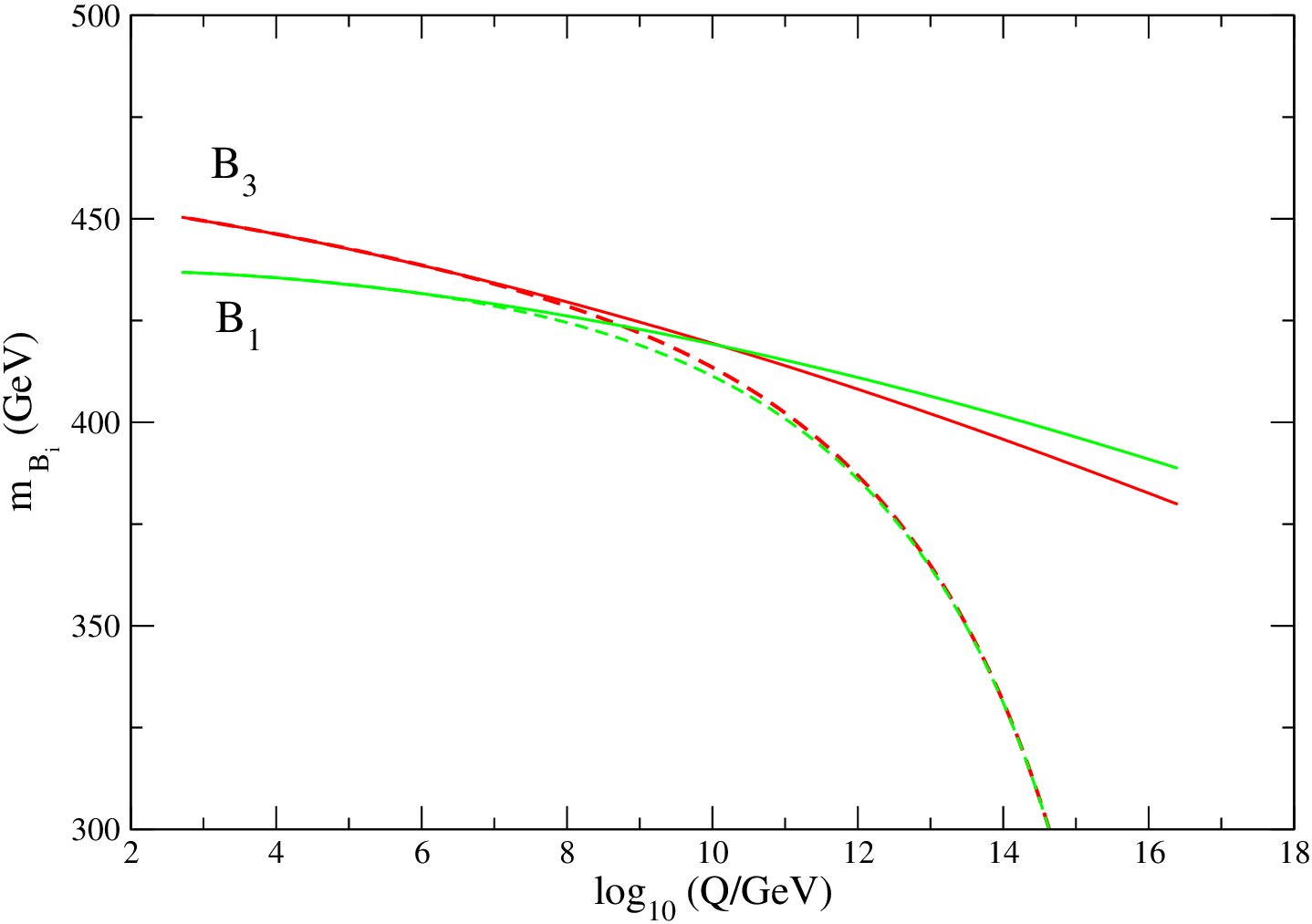}
\caption{\small Running of the soft scalar mass combinations
$m_{B_3}$ and $m_{B_1}$
with $\nfive = 3$ and $\tilde{\mu} = 10^{4}\,\gev$ for the 
mSUGRA input parameters $m_0 = 300\,\gev$, $m_{1/2} = 700\,\gev$, 
$\tan\beta = 10$, and $A_0 = 0$.  The dashed lines show the actual 
running of these parameters, while the solid lines show the
running from low to high using the RG equations of the MSSM,
ignoring the additional heavy multiplets.
The small deviations in these figures arise from higher loop effects.
} 
\label{fig:mb}
\end{center}
\end{figure*}  

\begin{figure*}[tbh]
\begin{center}
\vspace{1cm}
        \includegraphics[width = 0.65\textwidth]{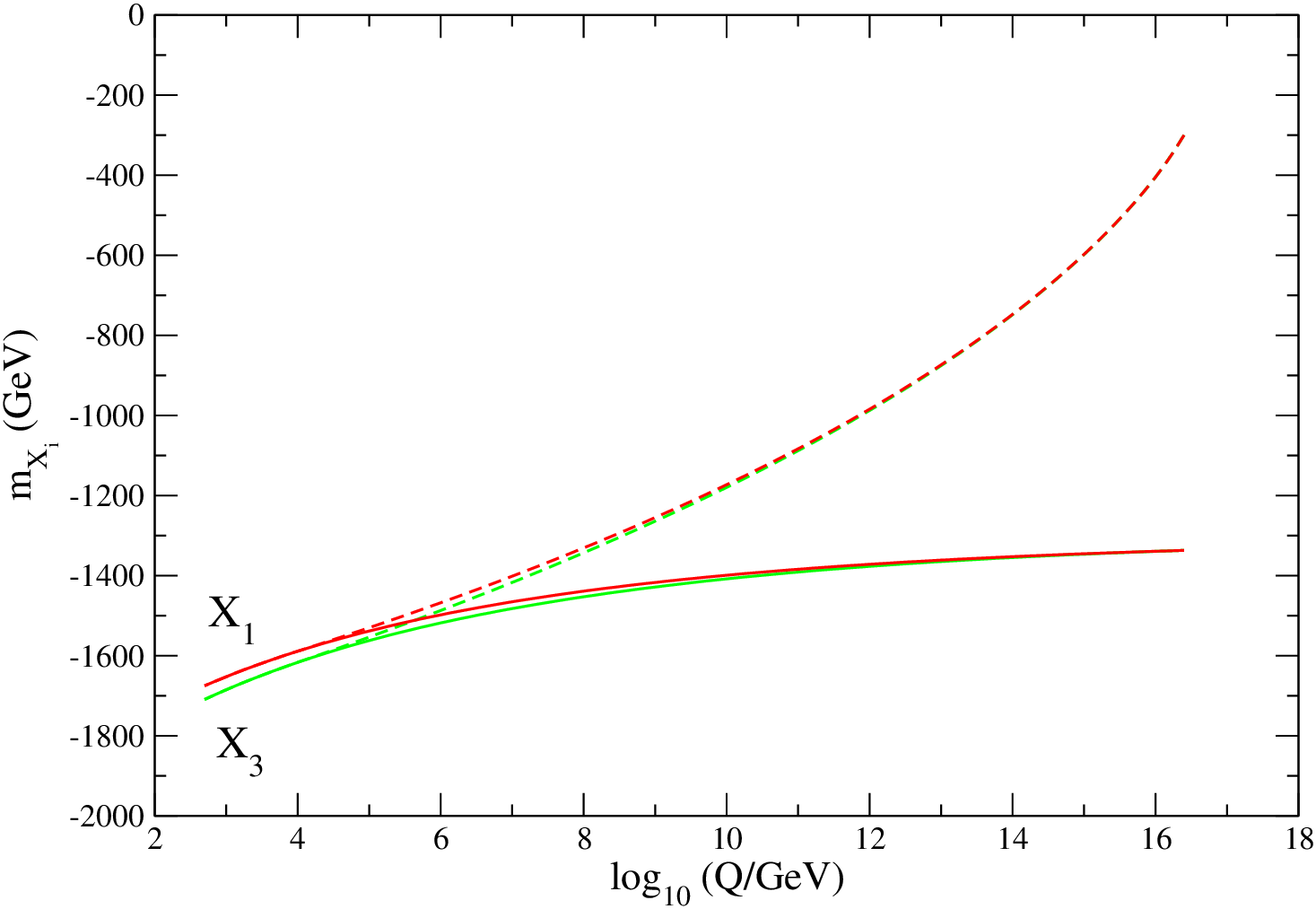}
\caption{\small Running of the soft scalar mass combinations 
$m_{X_3}$ and $m_{X_1}$
with $\nfive = 3$ and $\tilde{\mu} = 10^{4}\,\gev$ for the 
mSUGRA input parameters $m_0 = 300\,\gev$, $m_{1/2} = 700\,\gev$, 
$\tan\beta = 10$, and $A_0 = 0$.  The dashed lines show the actual 
running of these parameters, while the solid lines show the
running from low to high using the RG equations of the MSSM,
ignoring the additional heavy multiplets.  
The small deviations in these figures arise from higher loop effects.
} 
\label{fig:my}
\end{center}
\end{figure*}

  In Figs.~\ref{fig:mb} and \ref{fig:my} we show the running
of the mass combinations $m^{\phantom{2}}_{B_{1,3}}$ 
and $m^{\phantom{2}}_{X_{1,3}}$
(where $m_i = m_i^2/\sqrt{|m_i^2|}$) for $\nfive = 3$ and 
$\tilde{\mu} = 10^4\,\gev$ with the high scale mSUGRA input
values $m_0 = 300\,\gev$, $M_{1/2} = 700\,\gev$, $\tan\beta = 10$,
and $A_0=0$.  These figures also show the values of 
$m^{\phantom{2}}_{B_{1,3}}$ and $m^{\phantom{2}}_{X_{1,3}}$
that would be obtained by running up without
including the effects of the heavy new physics.  
Comparing these figures to Figs.~\ref{fig:univ1} and \ref{fig:univ2},
it is apparent that the splittings between $m_{B_1}^2$ and $m_{B_3}^2$,
and $m_{X_1}^2$ and $m_{X_3}^2$, are very much less than the high scale
splittings between the $Q$ and $U$ soft masses.  

  These relationship between the $B$ and $X$ soft mass combinations
from different families is a footprint left by
the full theory (including the heavy GUT multiplets) on the 
low-energy spectrum.  Since the scalar masses in the full theory are
universal at the high scale, the low scale splittings between the 
$B$ and $X$ soft mass combinations are very small.  On the other hand,
running the low scale $Q$ and $U$ scalar soft masses up within the MSSM
does not suggest any form of family universality among these masses.
Therefore, as we proposed above, the low-energy values of these
particular combinations of soft scalar masses can provide 
evidence for heavy new physics.\footnote{
We have checked that the small splittings between $m^{\phantom{2}}_{B_1}$ 
and $m^{\phantom{2}}_{B_3}$, 
as well as between $m^{\phantom{2}}_{X_1}$ and $m^{\phantom{2}}_{X_3}$,
arise from higher loop effects.}

  The analysis in this section shows that even though
new physics in the form of additional heavy GUT multiplets can 
significantly disrupt the predictions for the high scale soft 
spectrum obtained by running in the MSSM, 
certain key properties about the input spectrum can still
be deduced using collider scale measurements.  Most significantly,
the low-scale values of the gaugino masses and gauge couplings 
can be used to predict the approximate ratios of the high-scale 
values, provided gauge unification is preserved.  
  
  The effect of 
additional GUT multiplets on the scalar soft masses is more severe.
Extrapolating the soft masses without including the contributions
from the heavy GUT multiplets leads to a prediction for the input
soft masses that are generally too low.  The splittings between
the soft masses from different generations can be shifted as well.
Despite this, some of the flavor properties of the input soft
mass spectrum can be deduced by comparing the evolution of the mass
combinations in Eqs.~(\ref{massflav1}), (\ref{massflav2}),
and (\ref{massflav3}).  For example, $m_{B_3}^2 = m_{B_1}^2$
and $m_{L_1}^2 = m_{L_3}^2$ suggests some form of flavor 
universality (or an embedding in $SO(10)$), even if the scalar
masses extrapolated within the MSSM do not converge at $M_{GUT}$.
We expect that these special mass combinations could prove useful 
for studying other types of heavy new physics as well.


\section{The (S)Neutrino Connection\label{neut}}

  We have seen in the previous sections that taking the unification of
the gauge couplings as a serious theoretical input still leaves 
considerable room for experimental uncertainties and new physics 
to modify the extrapolated values of the soft supersymmetry breaking 
parameters at very high energies.
For example, a Fayet-Iliopoulos $D$-term for hypercharge
or additional complete GUT multiplets with intermediate scale masses
will not disrupt gauge coupling unification, but will in general change 
the running of the parameters of the model.  In this section we wish to
study the effect of additional intermediate scale singlet matter
with significantly large Yukawa couplings to the MSSM matter fields.
A particularly well-motivated example of this, and the one we consider,
are heavy singlet neutrino multiplets.

  The observed neutrino phenomenology can be accommodated by extending 
the matter content of the MSSM to include at least two right handed~(RH) 
neutrino supermultiplets that are singlets 
under the SM gauge group~\cite{Mohapatra:2005wg}.
Throughout the present work, we will assume there are 
three RH neutrino flavors.
The superpotential in the lepton sector is then given by
\bea
{\cal W}_l={\bf y_e}\ LH_dE+{\bf y_\nu}\ LH_uN_R - {1\over2}{\bf M_R} N_R N_R,
\eea
where $L,\,E$, and $N_R$ are respectively the $SU(2)_L$ doublet, 
$SU(2)_L$ singlet, and neutrino chiral supermultiplets, 
each coming in three families.  The quantities ${\bf y_{e}}$, 
${\bf y_{\nu}}$, and $\bf{M_R}$ are $3\times 3$ matrices 
in lepton family space. 
The $H_d$ and $H_u$ fields represent the usual Higgs multiplets.
The gauge-invariant interactions among leptons and
Higgs superfields are controlled by the family-space 
Yukawa matrices ${\bf y_e}$ and ${\bf y_\nu}$.
As is conventional, we shall implicitly work in a basis where
${\bf y_e}$ is diagonal.  Since the $N_R$ are singlets, we can also add
to the superpotential a Majorana mass ${\bf M_R}$ for these fields.

  Assuming the eigenvalues of ${\bf M_R}$ lie at a large 
intermediate mass scale, $10^9-10^{14}$ GeV, we can integrate 
out the RH neutrino superfields and obtain a term 
in the effective superpotential that leads to small
neutrino masses through the see-saw mechanism,
\beq
{\cal W }_{m_\nu}=-{1\over2}{\bf y_\nu^T M_R^{-1}y_\nu}\ LH_u LH_u.
\eeq
After electroweak symmetry breaking, the neutrino mass matrix becomes
\bea
{\bf(m_\nu)}={\bf y_\nu^T M_R^{-1}y_\nu}\, v_u^2,
\eea
where $v_u = \left<H_u\right>$.  The mass matrix can be
conveniently diagonalized by the transformation
\bea
{\bf(m_\nu^{diag})=U^T (m_\nu) U },
\eea
with $\bf{U}$ a unitary matrix.  This matrix $\bf{U}$ is
the usual PMNS matrix that describes lepton mixing relative to the flavor 
basis where the charged lepton Yukawa matrix ${\bf y_e}$ is diagonal.

  One can also write the neutrino Yukawa matrix as~\cite{Casas:2001sr}
\beq
{\bf y_\nu} = {1\over v_u} \bf \sqrt{M_R^{diag}}\ R\ \sqrt{m_\nu^{diag}}\
U^\dagger \label{ynucasas}
\eeq
where $v_u=\langle H_u\rangle$ and $R$ is a complex orthogonal matrix
that parametrizes our ignorance of the neutrino Yukawas.
As an estimate, Eq.~(\ref{ynucasas}) shows that the size of 
the neutrino Yukawa couplings will be on the order of
\beq
y_\nu \simeq \frac{0.57}{\sin\beta} 
\left({ M_R\over 10^{14} \gev}\right)^{1\over 2} 
\left({m_\nu\over 0.1 \eev}\right)^{1\over 2}.
\eeq
Thus, the neutrino Yukawa couplings can take large ${\cal O}$(1) values
comparable to the top Yukawa coupling for $M_R \sim 10^{14}\,\gev$.
If this is the case, then above the see-saw mass threshold the effects on 
the RG running of the MSSM soft parameters due to the neutrino Yukawas 
can be substantial. 

  The addition of RH neutrinos to the MSSM can lead to
lepton flavor violation~(LFV) through the RG running of the 
off-diagonal slepton mass terms~\cite{Borzumati:1986qx,Casas:1999tg}.
In this work we will only consider simple scenarios of 
neutrino phenomenology in which the amount of lepton flavor violation
induced by the heavy neutrino sector is small.
However, the observation of LFV signals could potentially provide
information about a heavy neutrino sector~\cite{Borzumati:1986qx,Casas:1999tg,
Deppisch:2002vz,Petcov:2005yh}.  
Precision measurements of the slepton
mass matrices can also be used to constrain possible heavy neutrino
sectors~\cite{Baer:2000hx,Davidson:2001zk}.
Heavy singlet neutrinos may also be related to the source of the baryon
asymmetry through the mechanism of leptogenesis~\cite{Buchmuller:2005eh}.

\subsection{Running Up}

  The strategy we use in this section is similar to the one followed 
in the previous sections.  We assume a universal high scale mass spectrum
at $M_{GUT}$, and RG evolve the model parameters down to the low scale
$M_{low} = 500\,\gev$ including the additional effects of 
the neutrino sector parameters.  The resulting low scale spectrum 
is then run back up to $M_{GUT}$ using the RG equations 
for the MSSM without including the neutrino sector contributions.  
As before, we use this procedure to illustrate 
the discrepancy between the extrapolated parameter values 
and their true values if the new physics
effects are not included in the running.

  To simplify the analysis, we make a few assumptions about
the parameters in the neutrino sector.  We choose the orthogonal
matrix $R$ to be purely real, and we take the heavy neutrino mass
matrix ${\bf M_R}$ to be proportional to the unit matrix,
${\bf M_R} = M_R\,\mathbb{I}$.  We also take the physical neutrino
masses to be degenerate.  This allows us to write
\beq
{\bf y_{\nu}} = \frac{0.57}{\sin\beta} 
\left({ M_R\over 10^{14}\,\gev}\right)^{1\over 2} 
\left({m_\nu\over 0.1\,\eev}\right)^{1\over 2}
{\bf R\,U}^{\dagger} := y_{\nu}\,{\bf R\,U}^{\dagger},
\eeq
so that the neutrino Yukawa couplings have the form of a universal
constant multiplying a unitary matrix.  To fix the value of $y_{\nu}$,
we will set $m_{\nu} = 0.1\,\eev$.\footnote{The diagonal neutrino 
mass matrix $\bf m_\nu^{diag}$ and the lepton mixing matrix 
$\bf U$ are measured at low scales, and one should really evaluate 
them at the intermediate scale $M_R$ by running the Yukawa 
couplings up~\cite{Petcov:2005yh}.  Since we are most interested in 
the effect of the neutrino Yukawas after reaching the intermediate scale, 
we will neglect this additional running below $M_R$.}.
This choice is close to being as large as possible while remaining 
consistent with the cosmological bounds on the sum of the neutrino 
masses, $\sum_{\nu} < 0.68\,\eev$~\cite{Spergel:2006hy}.  
Note that larger values of the neutrino masses tend to maximize 
the size of the resulting neutrino Yukawa couplings.

  We also need to impose boundary conditions on the RH sneutrino 
masses and trilinear couplings.  We take these new soft 
parameters to have universal and diagonal boundary conditions 
at the unification scale, 
\beq
m_{{N}_{ij}}^2 = m_0^2\,\delta_{ij},~~~~~{\rm and }~~~~~ a_{{\nu}_{ij}} 
= A_{0}\,y_{{\nu}_{ij}},
\label{nusoft}
\eeq
where $m_0$ and $A_0$ are the same universal soft scalar mass and 
trilinear coupling that we will apply to the MSSM fields in the 
analysis to follow.
With these choices for the neutrino parameters,
the effects of the neutrino sector on the (one-loop) RG running of the
MSSM soft terms take an especially simple form.
In particular, the amount of leptonic flavor mixing induced is expected
to be very small, and the diagonal and universal form of the 
neutrino sector soft terms, Eq.~(\ref{nusoft}), is approximately 
maintained at lower scales.

  To illustrate the effects of the neutrino sector, we set the 
high scale spectrum to coincide with the SPS-5 benchmark point, 
and we extend the corresponding soft terms to the neutrino sector.  
The input values at $M_{GUT}$ for this point are $m_0 = 150\,\gev$, 
$m_{1/2} = 300\,\gev$, $A_0 = -1000\,\gev$,
and $\tan\beta=5$ with $\mu$ positive.  These input values 
tend to magnify the effects of the neutrino sector because 
the large value of $A_0$ feeds into the running of the soft masses.  
Large neutrino Yukawa couplings alter the running 
of the top Yukawa coupling as well.

\begin{figure*}[tbh]
\begin{center}
\vspace{1cm}
        \includegraphics[width = 0.65\textwidth]{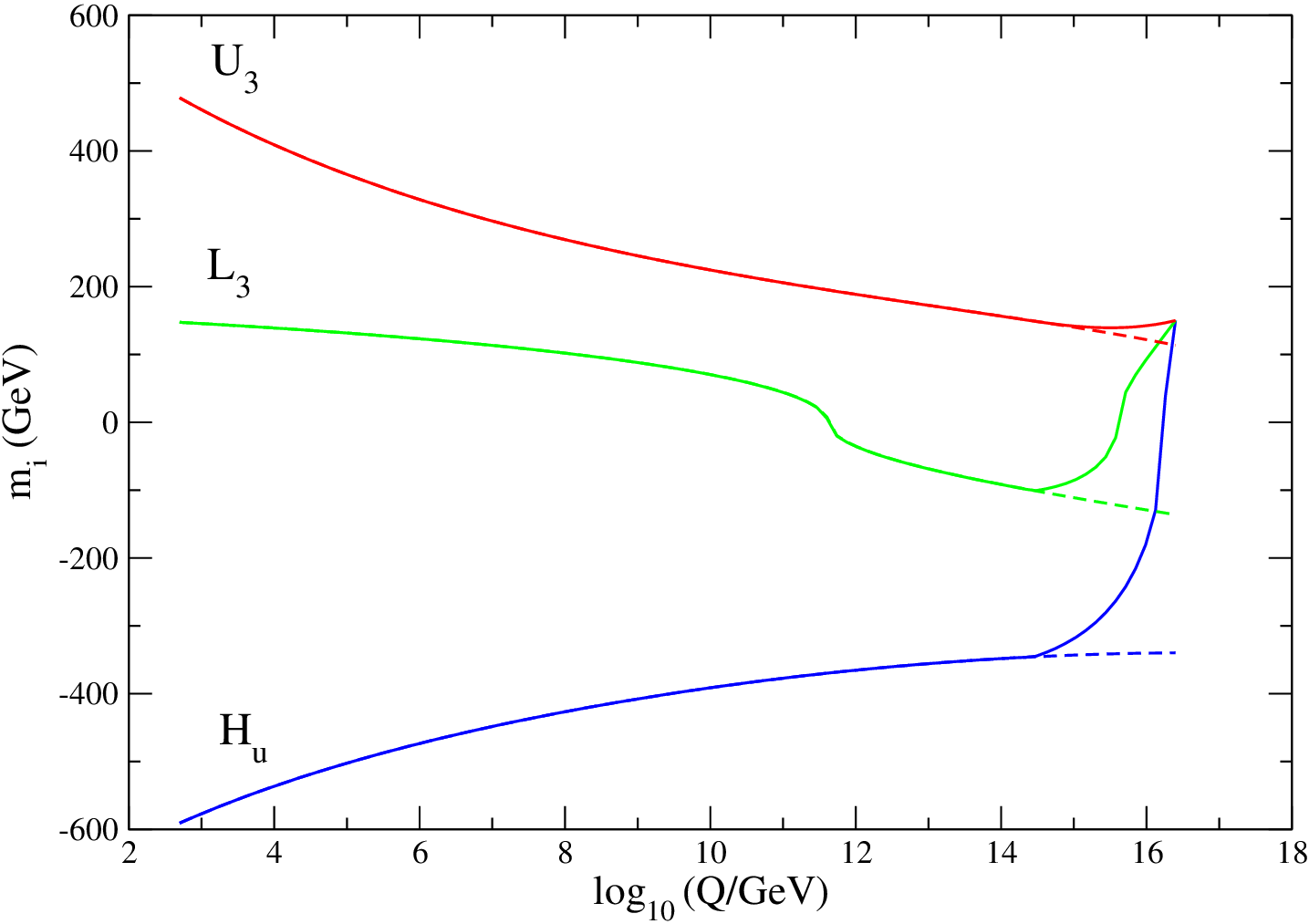}
\caption{\small Running of the soft scalar masses of $H_u$, $L_3$, 
and $U_3$ for SPS-5 input parameters at $M_{GUT}$ with three 
additional heavy RH neutrinos of mass $M_R = 10^{14}\,\gev$.  
The solid lines show the full running, including the neutrino 
sector effects, while the dashed lines show the low-to-high running
of the soft masses in the MSSM, with the neutrino sector effects omitted.
} 
\label{fig:yn-sps5-1}
\end{center}
\end{figure*}  

  Of the MSSM soft parameters, the greatest effects of the heavy
neutrino sector are seen in the soft scalar masses and the trilinear 
$A$ terms.  The gaugino masses are only slightly modified.
The evolution of the soft masses for the $H_u$, $L_3$, and $U_3$
fields from low to high are shown in Fig.~\ref{fig:yn-sps5-1}, both with 
and without including the effects of the neutrino sector for 
$M_R = 3\times 10^{14}\,\gev$.  
For this value of $M_R$ and with $\tan\beta = 5$, the Yukawa coupling 
is close to being as perturbatively large as possible.
The extrapolated values of $m_{H_u}^2$ and $m_{L_3}^3$ 
deviate significantly from the actual input values if the effects of the
neutrino sector are not taken into account in the RG evolution.
These fields are particularly affected because they couple directly 
to the heavy neutrino states through the neutrino Yukawa coupling.  
The shift in the running of $m_{U_3}^2$ arises indirectly
from the effect of the neutrino Yukawas on $m_{H_u}^2$ and the 
top Yukawa coupling $y_t$.

\begin{figure*}[tbh]
\begin{center}
\vspace{1cm}
        \includegraphics[width = 0.65\textwidth]{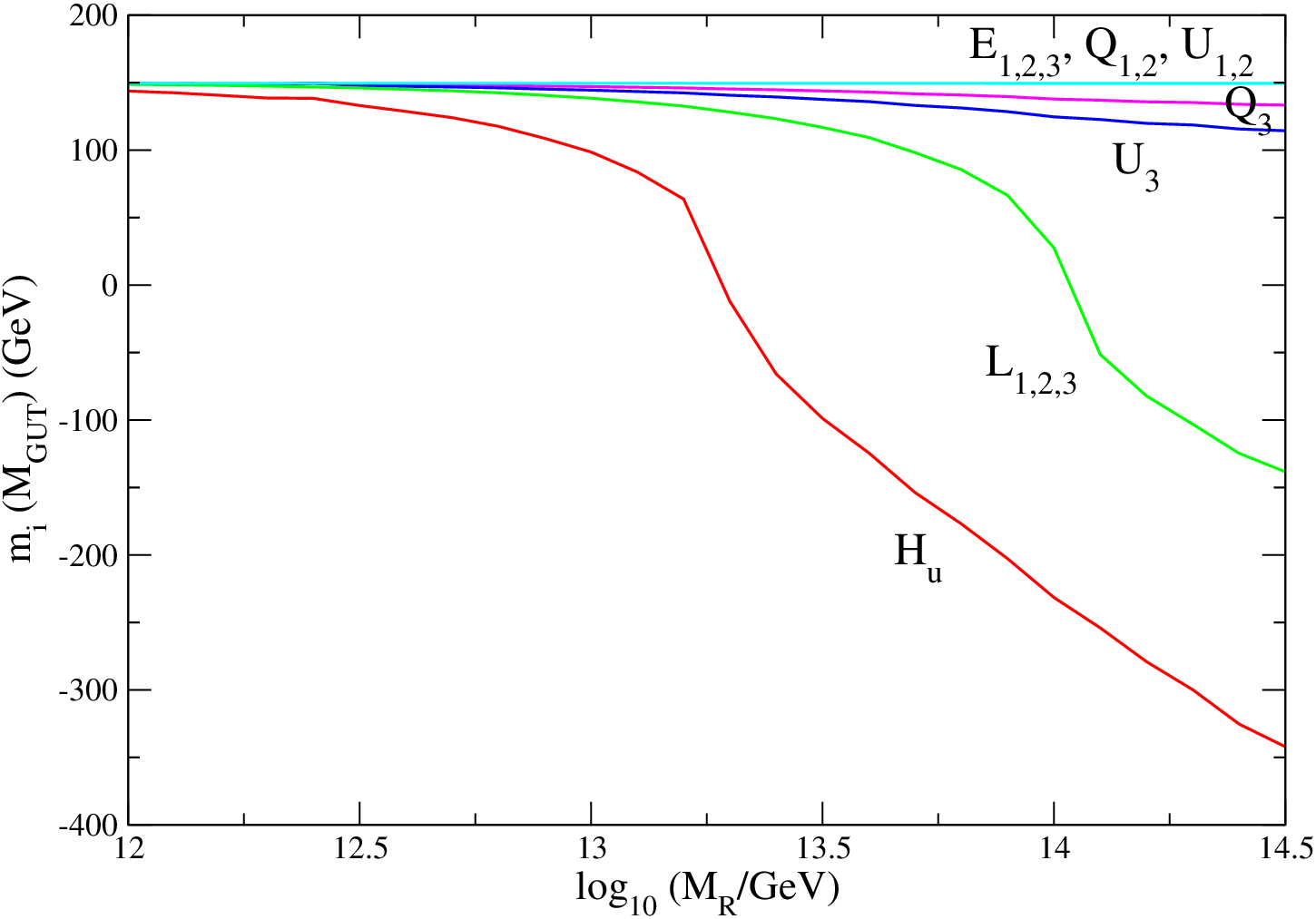}
\caption{\small The high scale ($M_{GUT}$) values of the soft scalar masses
extrapolated using the MSSM RG equations, without including neutrino
sector effects.  The low-scale ($500\,\gev$) values of the soft masses 
used in the extrapolation were obtained from SPS-5 input parameters 
at $M_{GUT}$ with three additional heavy RH neutrinos of mass 
$M_R$.  The deviations from $m_i(M_{GUT}) = 150\,\gev$ represent
the discrepancy between the MSSM extrapolated values and the correct
value in the full theory with heavy RH neutrinos.
These discrepancies are shown as a function of the heavy 
neutrino scale $M_R$.}
\label{fig:yn-sps5-2}
\end{center}
\end{figure*}

  In Fig.~\ref{fig:yn-sps5-2} we show the size of the discrepancies
in the extrapolated high scale values of a few of the soft scalar masses
if the neutrino sector effects are not included in the running.  
These discrepancies are plotted as a function of the heavy neutrino 
mass scale $M_R$.  As above, the high scale input spectrum 
consists of the SPS-5 values.  
Again, the soft masses $m_{H_u}^2$ and $m_{L_i}^2$ are 
altered the most, although the third generation squark soft masses 
also get shifted somewhat as a backreaction to the changes in $m_{H_u}^2$
and the top Yukawa coupling.  This plot also shows that the sizes
of the discrepancies remain quite small for $M_R$ less than $10^{13}$~GeV.
Values of $M_R$ considerably less than this are favored if leptogenesis
is to be the source of the baryon asymmetry~\cite{Buchmuller:2005eh}.
Neutrino masses well below $0.1\,\eev$ would also lead to less pronounced
deviations in the extrapolated soft masses.

  The deviations induced by not including the new neutrino sector
physics in the running take a similar form to those obtained by 
not taking account of heavy GUT multiplets.  For both cases,
the gaugino mass running is only modified in a very controlled way, 
while the soft scalar masses and trilinear
couplings deviate more unpredictably.  In particular, the high scale
flavor structure of the soft masses can be obscured.
With large neutrino Yukawa couplings, the third generation squark
masses receive additional contributions to their running relative
to the first and second generations due to the potentially large 
effect of the neutrino Yukawa couplings on $m_{H_u}^2$.
The sizes of these additional family-dependent 
shifts tend to be fairly small, as can be seen in 
Fig.~\ref{fig:yn-sps5-2}.  Furthermore, these effects cancel out 
in the mass combinations $m_{B_i}^2$ defined in Eqs.~(\ref{massflav1}) 
and (\ref{massflav2}), and we find $m_{B_1}^2 \simeq m_{B_3}^2$ 
at all scales, regardless of whether of not the neutrino effects are included.
On the other hand, the running of $m_{A_3}^2$ relative to $m_{A_1}^2$,
$m_{X_3}^2$ relative to $m_{X_1}^2$, and $m_{Y_3}^2$ relative 
to $m_{Y_1}^2$ need no longer coincide if there is a heavy
neutrino sector.  

  Finally, let us also mention that for more general neutrino sector 
parameters than those we have considered, there can arise significant 
lepton flavor mixing couplings in the MSSM slepton soft terms from 
the RG running~\cite{Borzumati:1986qx}.  Measurements of this mixing 
in lepton flavor violating processes can therefore provide an 
experimental probe of the heavy neutrino 
multiplets~\cite{Borzumati:1986qx,Casas:1999tg,Deppisch:2002vz,Petcov:2005yh}.  
A measured splitting among the three slepton masses $m^2_{L_i}$ would
also constitute another indication of the existence of a neutrino sector with
sizeable Yukawa couplings and nontrivial flavor structure. Both high and
intermediate energy data may be complementary and very useful in extrapolating
the MSSM soft terms to high energies.


\section{Putting it All Together: an Example\label{all}}

  In this section we summarize some of our previous results 
with an explicit example.  We begin with a low energy spectrum
of MSSM soft supersymmetry breaking parameters that we 
assume to have been measured at the LHC to an arbitrarily high precision.  
We then attempt to deduce the essential features of the underlying 
high scale structure by running the low energy parameters up 
and applying some of the techniques discussed in the previous sections.

  In our example we will make the following assumptions:
\begin{itemize}
\item The possible types of new physics beyond the MSSM
are:
\begin{itemize}
\item Complete $\five$ GUT multiplets with a common (SUSY)
mass scale $\tilde{\mu}$.  
\item Three families of heavy singlet (RH) neutrinos 
at the mass scale $M_R$.  
\item A fundamental hypercharge $D$ term.
\end{itemize}
In the case of complete GUT multiplets, we will assume further
that there are no superpotential interactions with the MSSM states
as in Section~\ref{gutmult}.
For heavy RH neutrinos, we will make the same set of assumptions
about the form of the mixing and mass matrices as in Section~\ref{neut}.
\item The high scale spectrum has the form of a minimal SUGRA model
(up to the scalar mass shifts due to a hypercharge $D$-term)
at the high scale $M_{GUT} \simeq 2.5\times 10^{16}\,\gev$.
\item This mSUGRA spectrum also applies to the soft parameters 
corresponding to any new physics sectors.  For example, a trilinear
$A$ term in the RH neutrino sector has the form ${\bf a}_{\nu}
= A_{0}\,{\bf y_{\nu}}$ at $M_{GUT}$, where $A_0$ is the universal
trilinear parameter.  
\end{itemize}

  These assumptions are not entirely realistic, but they make the 
analysis tractable.  Moreover, even though this exercise is 
highly simplified compared to what will be necessary should 
the LHC discover supersymmetry, we feel that it illustrates 
a number of useful techniques that could
be applied in more general situations.
With this set of assumptions, the underlying free parameters 
of the theory are: 
\beq
\begin{array}{cclcccc}
m_0,&m_{1/2},&A_0,~&
\xi,~&\nfive,&\tilde{\mu},~&M_R.
\end{array}
\eeq
where $m_0$, $m_{1/2}$ and $A_0$ are common mSUGRA inputs at $M_{GUT}$,
$\xi$ is the fundamental hypercharge $D$ term, $\nfive$ is the number
of additional $\five$ multiplets in the theory with a supersymmetric
mass $\tilde{\mu}$, and $M_R$ is the heavy neutrino scale.

\subsection{Step 1: Running Up in the MSSM}

  As a first step, we run the low energy spectrum 
up to the high scale $M_{GUT}$ using the RG equations
for the MSSM, without including any potential new physics effects.
The low energy MSSM soft spectrum we consider, 
defined at the low scale $M_{low} = 500\,\gev$, is given 
in Table~\ref{lowscalespectrum}.  In addition to these soft terms, 
we also assume that $\tan\beta = 7$ has been determined, 
and that the first and second generation soft scalar masses are equal.
With this set of soft terms, we have verified that the
low energy superpartner mass spectrum is 
phenomenologically acceptable using SuSpect~2.3.4~\cite{Djouadi:2002ze}.
The lightest Higgs boson mass is $114\,\gev$ for a top quark mass
of $m_t = 171.4\,\gev$.  

  Even before extrapolating the soft parameters, it is possible 
to see a number of interesting features in the spectrum.  
The most obvious is that the low scale gaugino masses have ratios 
close to $M_1:M_2:M_3 \simeq 1:2:6$.  
This suggests that the high scale gaugino masses have a universal 
value $m_{1/2}$, and provides further evidence for gauge unification.  
The low-energy value of the $S$ term, as defined in Eq.~(\ref{1:sterm}), 
is also non-zero and is in fact quite large, 
$S(M_{low}) \simeq (620\,\gev)^2$.  This indicates that there are
significant contributions to the effective hypercharge 
$D$ term in the high scale theory.  Since the $S$ term is non-zero, 
it is also not surprising that $S_{B-L} \simeq (446\,\gev)^2$, as defined
in Eq.~(\ref{1:sbl}), is non-zero as well.

\begin{table}[htb]
\vspace{0.3cm}
\center
\begin{tabular}{|c|c||c|}\hline
Soft Parameter &Low Scale Value&High Scale Value\\
&(GeV)&(GeV)\\ 
\hline
$M_1$ & 146&356\\
$M_2$ & 274&355\\
$M_3$ & 859&370\\ \hline\hline
$A_t$ & -956&-766\\
$A_b$ & -1755&-818 \\
$A_{\tau}$ & -737&-524\\ \hline\hline
$m_{H_u}$ & -700&419\\
$m_{H_d}$ & 350&236\\  \hline
$m_{Q_3}$ & 821& 549\\
$m_{U_3}$ & 603& 445\\
$m_{D_3}$ & 884& 501\\
$m_{L_3}$ & 356& 213\\
$m_{E_3}$ & 349& 404\\ \hline
$m_{Q_1}$ & 934& 532\\
$m_{U_1}$ & 872& 402\\
$m_{D_1}$ & 888& 501\\ 
$m_{L_1}$ & 357& 213\\
$m_{E_1}$ & 352& 404\\ \hline
\end{tabular}
\caption{\small The low-energy scale ($M_{low} = 500\,\gev$) 
soft supersymmetry breaking spectrum used in our analysis.  
The soft scalar masses listed in the table correspond to the 
signed square roots of the actual masses squared.
In this table we also the high scale values of these soft
parameters obtained by running them up to $M_{GUT} \simeq 
2.5\times 10^{16}\,\gev$ using the RG evolution appropriate
for the MSSM.
}
\label{lowscalespectrum}
\end{table}

\begin{figure*}[tbh]
\begin{center}
\vspace{1cm}
        \includegraphics[width = 0.65\textwidth]{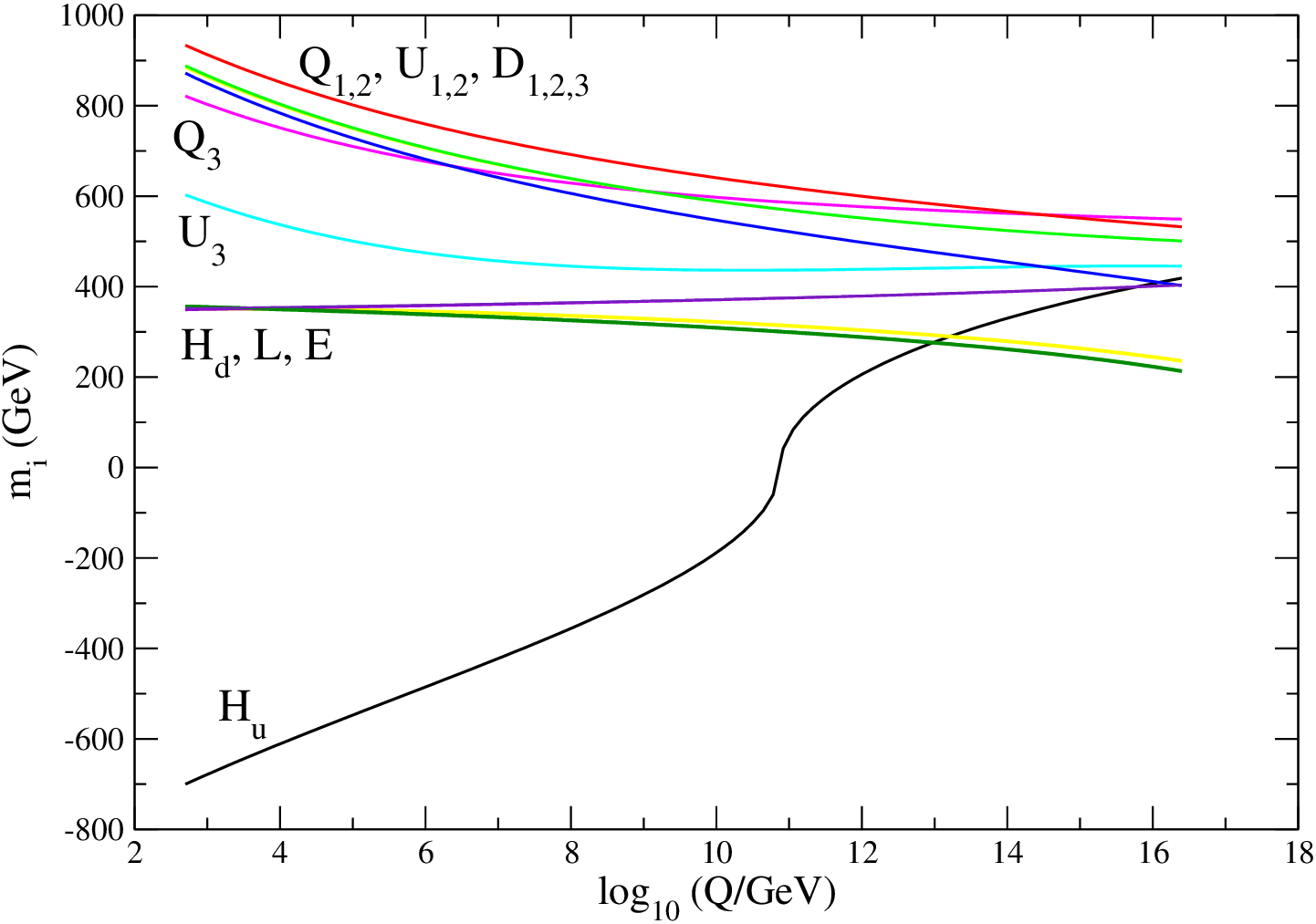}
\caption{\small Scale dependence of the soft scalar masses 
for the input soft parameters given in Table~\ref{lowscalespectrum}. 
No new physics effects beyond the MSSM were included in the running.
} 
\label{fig:explot1}
\end{center}
\end{figure*}

  The values of the soft parameters extrapolated to $M_{GUT}$
within the MSSM are listed in Table~\ref{lowscalespectrum}.
The MSSM running of the soft scalar masses is also shown 
in Fig.~\ref{fig:explot1}.  As anticipated, the gaugino masses 
unify approximately to a value 
$M_1\simeq M_2\simeq M_3\simeq m_{1/2} = 350\,\gev$
at $M_{GUT}$.  The high scale pattern of the soft scalar masses
(and the trilinear $A$ terms) shows less structure, and is clearly 
inconsistent with mSUGRA high scale input values.  

  Since $S(M_{low})$ is large and non-zero, 
we are motivated to look for a hypercharge $D$ term contribution
to the soft scalar masses.  
Such a contribution would cancel in the mass combinations
\beq
\Delta m^2_{ij}=(Y_j\,m_i^2-Y_{i}\,m_j^2)/(Y_j-Y_i).\label{hypermasscomb}
\eeq
If the high scale soft masses have the form
$m_i^2 = m_0^2 + \sqrt{\frac{3}{5}}g_1\,Y_i\,\xi$, 
as in mSUGRA with a hypercharge $D$ term,
these combinations will all be equal to $m_0^2$ at this scale.  
The high scale soft masses here, extrapolated within the MSSM, exhibit 
no such relationship.  Even so, these mass combinations 
will prove useful in the analysis to follow.

  It is also interesting to compare the pairs of mass combinations 
$m_{A_i}^2,\,m_{B_i}^2,\,m_{X_i}^2$, and 
$m_{Y_i}^2$ for $i=1,\,3$, as defined in Section~\ref{gutmult}.
Of these, the most useful pair is $m_{B_1}^2$ and $m_{B_3}^2$.
At the low and high scales (extrapolating in the MSSM) this
pair obtains the values
\beq
\begin{array}{cccccc}
m_{B_1}^2(M_{low}) &\simeq& (441\,\gev)^2,~~~&
m_{B_3}^2(M_{low}) &\simeq& (452\,\gev)^2,\\
m_{B_1}^2(M_{GUT}) &\simeq& (392\,\gev)^2,~~~&
m_{B_3}^2(M_{GUT}) &\simeq& (392\,\gev)^2.
\end{array}
\eeq
The near equality of $m_{B_1}^2$ and $m_{B_3}^2$ at the high
scale is particularly striking.  By comparison, there is a
significant inter-family splitting that occurs
between the high scale soft masses of $U_1$ and $U_3$,
\beq
m_{U_1}^2(M_{GUT}) = (402\,\gev)^2,~~~~~m_{U_3}^2(M_{GUT}) = (445\,\gev)^2.
\eeq
This apparent fine-tuning among the soft masses that make up 
the $B$-type combinations is suggestive of an underlying structure
in the theory.  However, based on the values of the individual 
soft masses extrapolated within the MSSM, such a structure 
is not obvious.  Instead, we can interpret this as a hint 
for new intermediate scale physics.

  Without our guiding assumptions, the high scale spectrum listed
in Table~\ref{lowscalespectrum} obtained by running up in the MSSM
does not exhibit any particularly remarkable features aside
from the universality of the gaugino masses. 
Even so, the curious relationship between
the $m_{B_1}^2$ and $m_{B_3}^2$ mass combinations provides a strong
hint that we are missing something.  It is not clear how strong this
hint would have been 
had we also included reasonable uncertainties in the 
low scale parameter values.

\subsection{Step 2: Adding GUT Multiplets}
 
  As a first attempt to fit the low energy soft spectrum to the
class of models outlined above, let us consider adding additional 
vector-like GUT multiplets the the theory at the scale $\tilde{\mu}$.
We try this first because, as we found in Sections~\ref{gutmult}
and \ref{neut}, the contributions from such multiplets are potentially
much larger than those due to heavy singlet neutrinos.

  In adding the new GUT multiplets, we will make use of our starting
assumptions about the possible forms of new physics.  Given the large
value of $S(M_{low})$, there appears to be significant hypercharge
$D$ term.  Also from our assumptions, this $D$ term will contribute
to the soft scalar masses of the heavy GUT multiplets, which will
in term feed into the running of the MSSM scalar masses through the
$S$ term above the scale $\tilde{\mu}$.
For this reason, it is safer to work with the mass differences
defined in eq.~(\ref{hypermasscomb}) whose running (to one-loop) does not
depend on the S-term.
 
  Among the low scale soft masses listed in  Table~\ref{lowscalespectrum}, 
we expect the slepton soft mass $m^2_{E_1}$ to be among the easiest
to measure, and the least susceptible to new physics effects.
Thus, we will use it as a reference mass in all but two of
the differences we choose. The mass differences
we consider are
\beq
\begin{array}{cccc}
\Delta m^2_{Q_1E_1},&\Delta m^2_{U_1E_1},&
\Delta m^2_{D_1E_1},&\Delta m^2_{L_1E_1}\\
\Delta m^2_{H_dE_1},&\Delta m^2_{H_uE_1},&
\Delta m^2_{Q_1D_1},&\Delta m^2_{Q_3D_1},\\
\Delta m^2_{Q_3E_1},&\Delta m^2_{U_3E_1},&
\Delta m^2_{D_3E_1},&\Delta m^2_{L_3E_1}.
\end{array}
\label{hypermasscombbasis}
\eeq
These depend on several independent mass measurements.

\begin{figure*}[tbh]
\begin{center}
\vspace{1.2cm}
        \includegraphics[width = 0.85\textwidth]{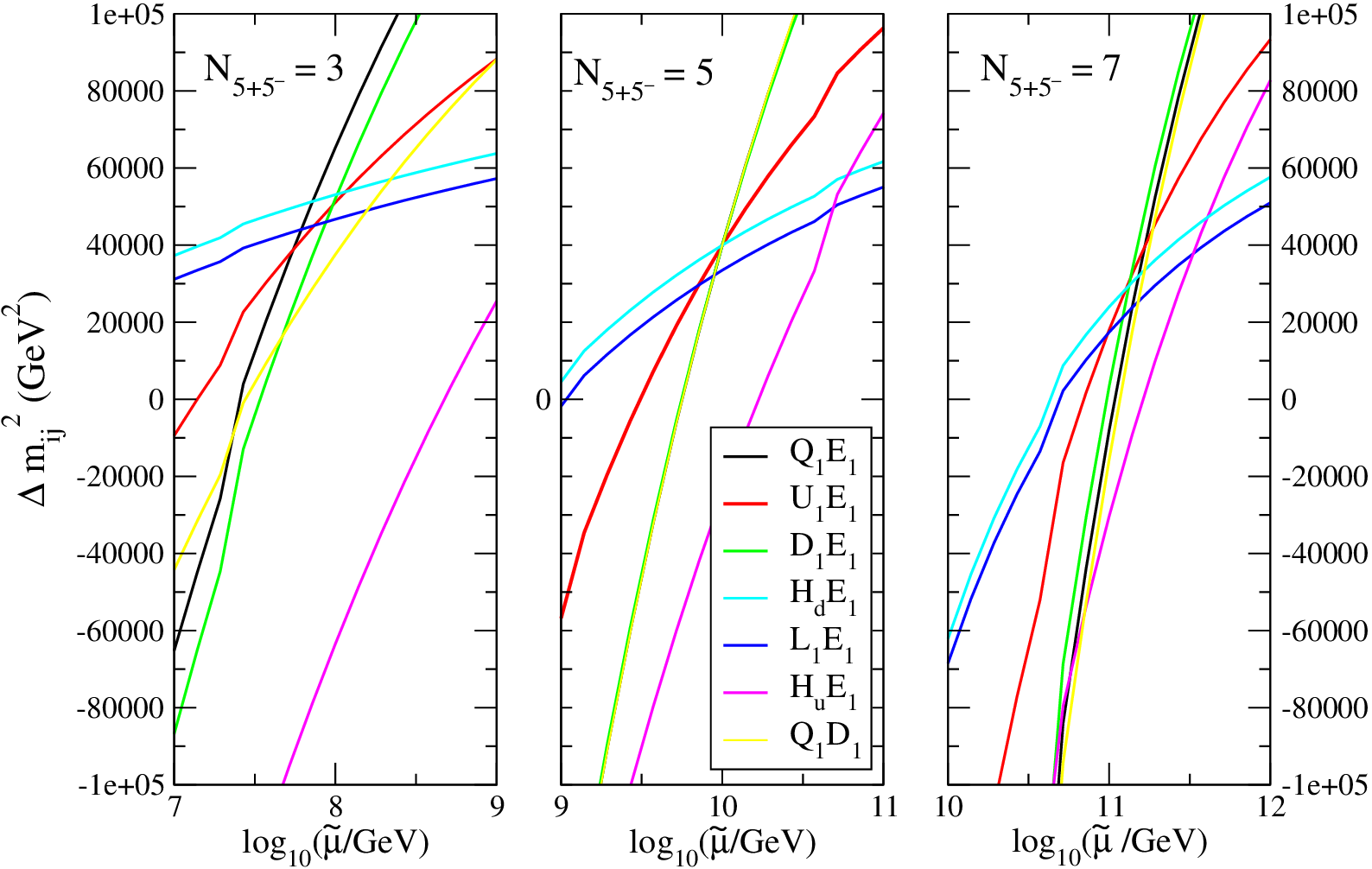}
\caption{\small High scale values for the soft scalar mass 
differences $\Delta{m}_{ij}^2$ defined in Eq.~(\ref{hypermasscombbasis}) 
as a function of the mass scale $\tilde{\mu}$ of the $\nfive$ 
heavy GUT multiplets for $\nfive = 3,\,5,\,7$.
} 
\label{fig:n5scan}
\end{center}
\end{figure*}

  In Fig.~\ref{fig:n5scan} we show the high scale values 
of these mass differences obtained by running up the low scale soft masses 
while including a given number of $\nfive$ additional $\five$ GUT multiplets
at the scale $\tilde{\mu}$.  For each of the plots, nearly all
the mass differences unify approximately, as they would be expected
to do if the underlying theory has a mSUGRA spectrum.
The best agreement with a mSUGRA model is obtained for 
$\nfive = 5$ with $\tilde{\mu} = 10^{10}\,\gev$. (More precisely, 
the agreement is obtained when a shift $\Delta b=-5$  is applied to 
the gauge beta function coefficients $b_i$ at the scale 
$\tilde{\mu} = 10^{10}\,\gev$).

  Taking $\nfive = 5$ and $\tilde{\mu} = 10^{10}\,\gev$,
the high scale values of the mass differences are 
\bea
&&\Delta{m}_{Q_1E_1}^2 ~~~=~~ (200\,\gev)^2 ~=~ 
\Delta{m}_{U_1E_1}^2 ~=~ \Delta{m}_{D_1E_1}^2
~=~\Delta{m}_{H_dE_1}^2\\
&&\begin{array}{cclccl}
\Delta{m}_{Q_3E_1}^2 &=& (197\,\gev)^2,&
\Delta{m}_{D_3E_1}^2 &=&(200\,\gev)^2\\
\Delta{m}_{H_uE_1}^2 &=& -(157\,\gev)^2,&
\Delta{m}_{L_{1,3}E_1}^2 &=& (183\,\gev)^2\\
\end{array}\nnmb
\eea
Most of these high scale values coincide, 
suggesting a mSUGRA value for the universal soft
scalar mass of about $m_0 = 200\,\gev$.  On the other hand,
the soft mass differences involving the $H_u$ and $L$ fields show a 
significant deviation from this near-universal value.
Based on the results of Section~\ref{neut}, these are precisely 
the scalar masses that are most sensitive to a heavy singlet 
neutrino sector.

  Using this same choice of new physics parameters, 
we can also estimate the values of the 
other mSUGRA parameters.  Heavy singlet neutrinos are not expected 
to significantly alter the running of the gaugino soft mass parameters.
If we run these up to $M_{GUT}$ including $\nfive=5$ 
additional GUT multiplets at $\tilde{\mu}=10^{10}\,\gev$, 
we find $m_{1/2} \simeq 700\,\gev$.
Doing the same for the trilinear couplings,
we do not find a unified high scale value for them.
Instead, we obtain $A_t(M_{GUT}) = -401\,\gev$, $A_{\tau} = -407\,\gev$,
and $A_{b} = -500\,\gev$.  This is not surprising since a heavy
RH neutrino sector would be expected to primarily modify
$A_t$ and $A_{\tau}$, while having very little effect on $A_b$.  
Thus, we also expect $A_0 \simeq -500\,\gev$. 

  It is possible to estimate the value of the hypercharge $D$ term as well.  
Using the hypothesis $m_i^2 = m_0^2 + \sqrt{\frac{3}{5}}g_1\,Y_i\xi$
at the high scale, we find
\beq
\xi(M_{GUT}) 
= \sqrt{\frac{5}{3}}\frac{1}{g_1}\,{\left(m_{E_1}^2-m_{H_d}^2\right)}/
{(Y_E-Y_{H_d})}\simeq (494\,\gev)^2.\label{xivalue}
\eeq
We obtain similar values from the corresponding combinations of other
mass pairs with the exception of $L$ and $H_u$.  Based on our previous
findings, we suspect that the $L$ and $H_u$ soft masses are modified
by a heavy RH neutrino sector.  

  Note that had we included experimental and theoretical uncertainties
it would have been considerably more difficult to distinguish
different values of $\nfive$ and $\tilde{\mu}$.
Instead of finding a single value for $\five$ and a precise value
for $\tilde{\mu}$, it is likely that we would have only been
able to confine $\nfive$ and $\tilde{\mu}$ to within finite ranges.

\subsection{Step 3: Adding a Heavy Neutrino Sector}

  By adding $\nfive = 5$ complete $\five$ multiplets at the 
scale $\tilde{\mu} = 10^{10}\,\gev$ and a hypercharge $D$ term,
we are nearly able to fit the low scale spectrum given in
Table~\ref{lowscalespectrum} to a mSUGRA model with 
$m_0 = 200\,\gev$, $m_{1/2} = 700\,\gev$, and $A_0 = -500\,\gev$.
However, there are several small deviations from this picture,
most notably in the soft masses for $H_u$ and $L$ as well as
the trilinear couplings $A_t$ and $A_{\tau}$.  We attempt to 
fix these remaining discrepancies by including heavy RH neutrinos
at the scale $M_R$.  

\begin{figure*}[tbh]
\begin{center}
\vspace{1.2cm}
        \includegraphics[width = 0.65\textwidth]{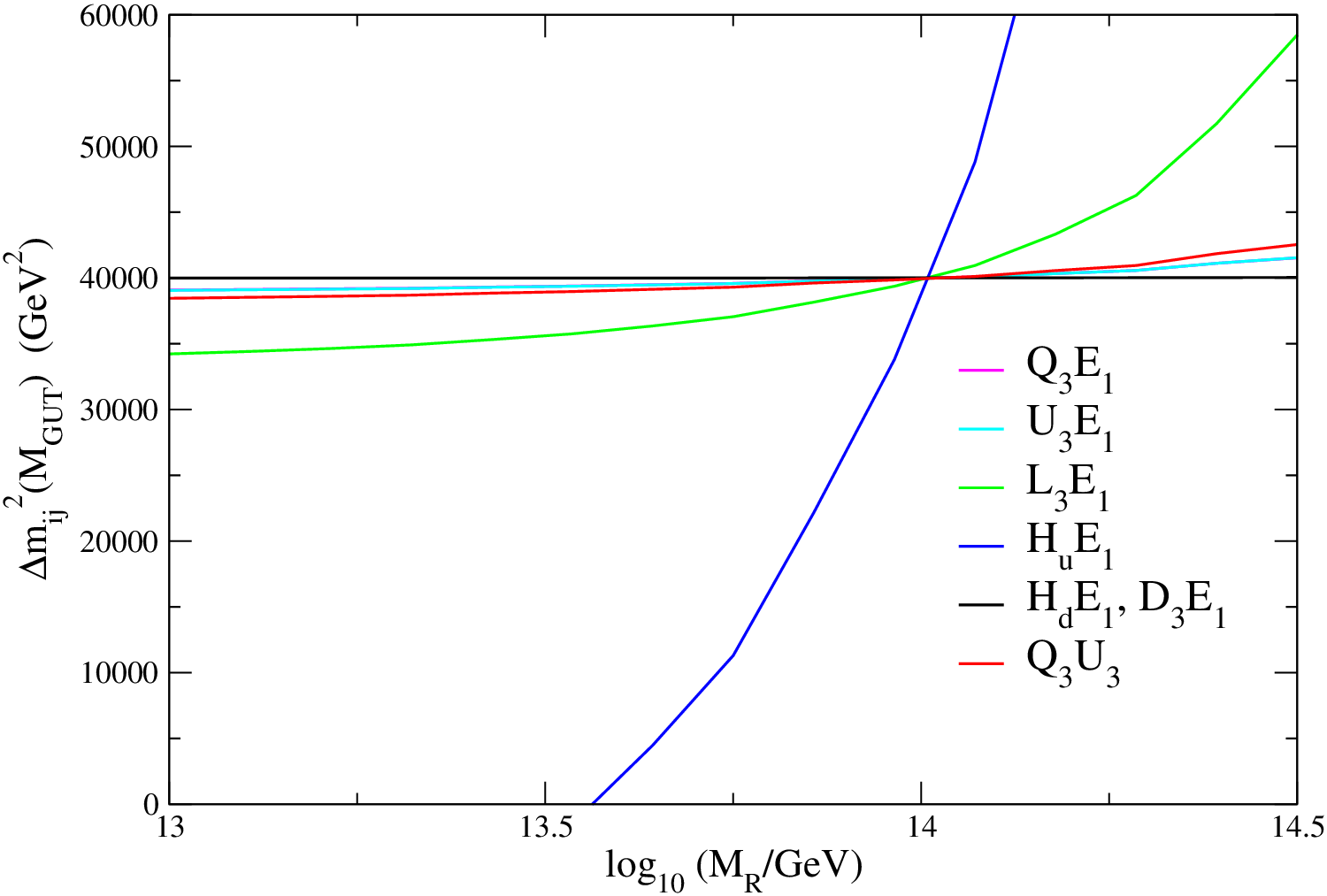}
\caption{\small High scale values for third family and Higgs
boson soft scalar mass differences $\Delta{m}_{ij}^2$ 
(as defined in Eq.~(\ref{hypermasscomb})) as a function
of the mass scale $M_R$ of the heavy RH neutrinos.
$\nfive = 5$ GUT multiplets were included in the running
above the scale $\tilde{\mu} = 10^{10}\,\gev$.
} 
\label{fig:mrscan}
\end{center}
\end{figure*}

  Given our initial assumptions about the form of a possible 
RH neutrino sector, the only independent parameter in this sector
is the heavy mass scale $M_R$.  To investigate the effects of
RH neutrinos, we examine the high scale values of the 
mass differences given in Eq.~(\ref{hypermasscomb}) 
for the third generation scalars and $H_u$,
using $m^2_{E_1}$ as a reference mass.
We add a RH neutrino sector with heavy mass $M_R$
and run the low scale parameters listed in  Table~\ref{lowscalespectrum}
subject to the additional neutrino effects, as well as those from
$\nfive = 5$ heavy GUT multiplets with $\tilde{\mu} = 10^10\,\gev$.  
The result is shown in Fig.~\ref{fig:mrscan}, in which we scan over $M_R$.  
This plot shows that if we include heavy
RH neutrinos at the mass scale near $M_R = 10^{14}\,\gev$, 
all the mass differences will flow to a universal value of about
$m_0 = 200\,\gev$ at the high scale.  

\begin{figure*}[tbh]
\begin{center}
\vspace{1.2cm}
        \includegraphics[width = 0.65\textwidth]{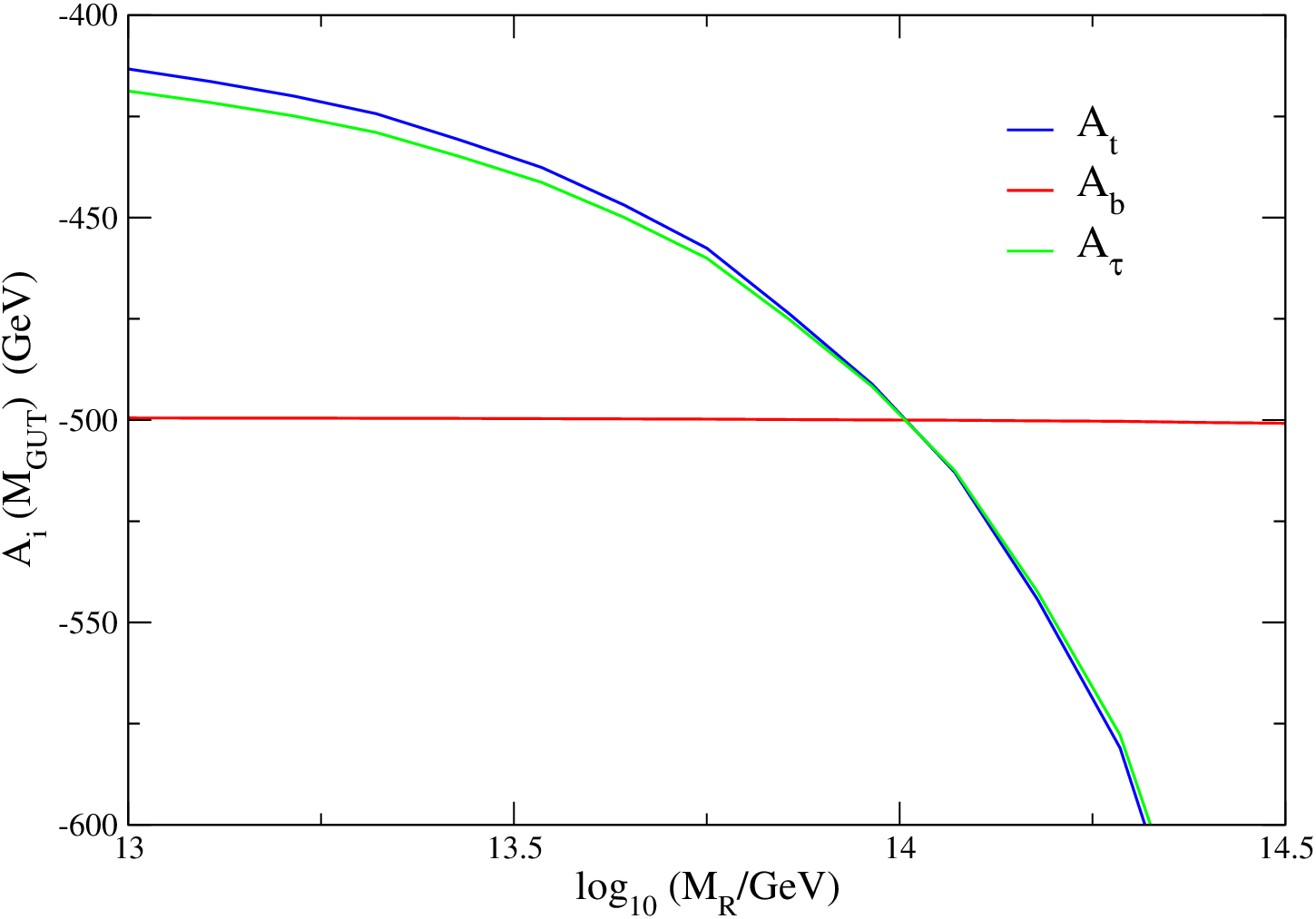}
\caption{\small High scale values for the soft trilinear
$A$ terms as a function of the mass scale of the heavy RH neutrinos.
$\nfive = 5$ GUT multiplets were included in the running
above the scale $\tilde{\mu} = 10^{10}\,\gev$.
} 
\label{fig:mrascan}
\end{center}
\end{figure*}

  We can further confirm this result by examining the 
high scale trilinear couplings obtained by this procedure.
These also attain a universal value, $A_0 = -500\,\gev$, at the high scale
for $M_R =10^{14}\,\gev$ as shown in Figure \ref{fig:mrascan}.  
These universal values are consistent with those we hypothesized before 
the inclusion of heavy neutrino sector effects.
Similarly, we can also study the value of $\xi$ obtained using
Eq.~(\ref{xivalue}), but using the high scale $L$ and $H_u$ soft masses 
computed by including heavy RH neutrinos in their RG evolution.
As before, we obtain a value $\xi \simeq (494\,\gev)^2$.

\subsection{Summary}

We have succeeded in deducing a high scale mSUGRA model 
augmented by heavy new physics effects
that reproduces the soft spectrum in  
Table~\ref{lowscalespectrum}.
The relevant parameter values are:
\beq
\begin{array}{cclcclccl}
m_0 &=& 200\,\gev,~~~&m_{1/2} &=& 700\,\gev,&A_0&=&-500\,\gev,\\
\nfive &=& 5,&\tilde{\mu} &=& 10^{10}\,\gev,\\
M_R &=& 10^{14}\,\gev,&\xi &=& (494\,\gev)^2.
\end{array}
\eeq
Our example did not include potential uncertainties in the input 
soft parameter values.  It is likely
that such uncertainties would make the analysis more challenging.

  Running the low scale soft parameters up to $M_{GUT}$ within the MSSM,
and without including any new physics, we obtained a reasonable but
mostly unremarkable high scale soft spectrum.  The most obvious feature
of this spectrum is the unification of the gaugino masses.  
A more subtle aspect of the high scale spectrum is the small splitting
between $m_{B_1}^2$ and $m_{B_3}^2$ relative to that between the 
$m_{Q_1}^2$ and $m_{Q_3}$, and $m_{U_1}^2$ and $m_{U_3}^2$.  
This feature hinted at an underlying family-universal 
flavor spectrum obscured by new physics effects.  
It is not clear whether this hint would survive in a more 
complete treatment that included uncertainties in the
input parameter values.  By adding new physics, 
in the form of heavy GUT multiplets and RH neutrinos, 
a simple mSUGRA structure emerged.

\section{Conclusions\label{conc}}

  If supersymmetry is discovered at the LHC, the primary challenge
in theoretical particle physics will be to deduce the source 
of supersymmetry breaking.  By doing so, we may perhaps learn about 
the more fundamental theory underlying this source.  In most models 
of supersymmetry breaking, the relevant dynamics take place at energies 
much larger than those that will be directly probed by the LHC.  
It is therefore likely that the soft supersymmetry
breaking parameters measured by experiment will have to be extrapolated
to higher scales using the renormalization group.  Given the apparent
unification of gauge couplings in the MSSM only slightly below
$M_{\rm Pl}$, we may hope that there is little to no new physics 
between the LHC scale and the supersymmetry breaking scale so that
such an extrapolation can be performed in a straightforward way.

  Gauge unification still allows for some types of new
physics at intermediate scales such as complete GUT multiplets
and gauge singlets.  If this new physics is present, 
RG evolving the MSSM soft parameters without including the new physics
effects can lead to an incorrect spectrum of soft parameters 
at the high scale.  Even without new intermediate physics, 
if some of the MSSM soft parameters are only poorly determined 
at the LHC, or not measured at all, there can arise significant
uncertainties in the RG running of the soft masses that have
been discovered.

  In the present work we have investigated both of these 
potential obstacles to running up in the MSSM.  The soft scalar 
masses are particularly sensitive to these effects, 
but we find that the gaugino soft masses, 
and their ratios in particular, are considerably more robust.  
If any one of the scalar soft masses goes unmeasured at the LHC, 
the running of the remaining these soft terms 
can be significantly modified by the effects of the hypercharge $S$ term.
These effects are especially severe for the slepton soft masses,
which otherwise do not tend to run very strongly at all.  
We find that the uncertainties due to the $S$ term can be avoided if
we consider the soft mass differences
$(Y_j\,m_i^2 - Y_i\,m_j^2)$, where $Y_i$ denotes the hypercharge of
the corresponding field.  If all the soft mass are measured, 
the soft scalar mass combinations $S$ and $S_{B-L}$, defined
in Eqs.~(\ref{1:sterm}) and (\ref{1:sbl}), 
provide useful information about potential 
GUT embeddings of the theory.

  We have also investigated the effects of two plausible types 
of new physics beyond the MSSM that preserve consistency 
with gauge unification; namely complete vector-like 
GUT multiplets and heavy singlet neutrinos.   
In the case of complete GUT multiplets, 
extrapolating the measured low-energy soft parameter values 
without including the additional charged matter in
the RG running leads to high scale predictions for the gaugino masses 
that are too small, and soft scalar masses that are too large.  
Even so, the ratios of the gaugino masses at the high scale 
are not modified at leading order, and can be predicted from 
the low-energy measured values provided gauge unification occurs.  
The extrapolated values of the scalar masses
are shifted in more complicated ways, and relationships such as family
universality at the high-scale can be obscured.  Despite this, certain
hints about the underlying flavor structure of the soft masses 
can still be deduced from the properties of special linear combinations
of the soft masses, such as $(2\,m_{Q_3}^2-m_{U_3}^2-m_{D_3}^2)$ relative
to $(2\,m_{Q_1}^2-m_{U_1}^2-m_{D_1}^2)$.  

  The running of the MSSM soft masses can also be modified if
there are heavy singlet neutrino chiral multiplets in the theory.
These can induce small masses for the standard model neutrinos
through the see-saw mechanism.
If the singlet neutrino scale is very heavy,
greater than about $10^{13}\,\gev$, the corresponding neutrino
Yukawa couplings can be large enough to have a significant effect
on the running of the soft masses of $H_u$ and $L$.  We have studied
the size of these effects, as well as the shifts in the
other soft masses.  The extrapolated values of the gaugino masses
and the squark soft masses are only weakly modified by heavy
neutrino sector effects.  

  In Section~\ref{all} we applied the methods described above
to a specific example.  In this example, the scalar masses 
have a common high scale value, up to a hypercharge $D$ term.
However, because of the presence of heavy new physics, this simple
structure does not emerge when the low scale soft scalar masses 
are extrapolated up to $M_{GUT}$ using the RG equations of the MSSM.
Based on the low-energy spectrum alone, we were able to deduce
the presence of the hypercharge $D$ term.  The presence of additional
new physics was suggested by the fact that the
splitting between $m_{U_3}^2$ and $m_{U_1}^2$ was considerably
larger than the related splitting between $m_{B_3}^2$ and $m_{B_1}^2$.
By studying the scalar mass combinations of Eq.~(\ref{hypermasscomb}) 
and including heavy GUT multiplets and a right-handed neutrino sector,
we were able to reproduce the low-energy soft spectrum with an 
underlying mSUGRA model.  

  One can also invert this perspective of 
overcoming new physics obstacles, and instead view these 
obstacles as providing information and opportunities.  
As the example of Section~\ref{all} illustrates, 
analyses of the kind we consider here probe new physics 
in indirect ways that can lead to convincing arguments for its 
existence or absence.  Such analyses may be the main way
to learn about new physics that cannot be studied directly.

  If the LHC discovers new physics beyond the standard model,
it will be a challenge to extract the Lagrangian parameters from the data.  
It may also be difficult to correctly extrapolate these parameters 
to higher scales in order to deduce the underlying theory that gives
rise to the low energy Lagrangian.  In the present work 
we have begun to study this second aspect of the 
so-called LHC inverse problem, and we have found 
a few techniques to address some of the potential obstacles.  
However, our study is only a beginning.  We expect that a number
of additional techniques for running up could be discovered with more
work.  A similar set of techniques could also be applied to 
understanding the high scale origin of other types of new physics
beyond the standard model.  These techniques deserve further study.

\section*{Acknowledgements}

We would like to thank Aaron Pierce, Lian-Tao Wang, and James Wells
for helpful discussions and useful comments on the manuscript.
This work is supported by the Department of Energy, and the Michigan
Center for Theoretical Physics.


\appendix

\section*{Appendix: Useful Combinations of Scalar Masses}

  In this appendix we collect some combinations of 
soft scalar masses that are particularly useful for running up.

\subsection*{$S$ Term Effects}

  The mass differences
\beq
\Delta m_{ij}^2 = (Y_j\,m_i^2 - Y_i\,m_j^2)/(Y_j-Y_i)
\label{stermmass}
\eeq
are useful when there is a non-vanishing hypercharge $D$ term.
A hypercharge $D$ term can shift the low energy values 
of the soft scalar masses, and can also modify their RG 
running through the $S$ term, as discussed in Section~\ref{sterm}, 
and in Refs.~\cite{dterm,dgut,stermrefs}.  
The effects of a hypercharge $D$ term 
cancel out these mass differences, as well as in the RG equations for them.

  This feature is helpful for running up because the effect 
of the $D$ term on the running is determined by the low
scale value of the $S$-term, which depends on all the soft scalar
masses (of hypercharged fields) in the theory.  
If one of these soft masses goes unmeasured, 
there will be a large uncertainty in the value of the $S$ term, 
and this in turn will induce a significant uncertainty in the high 
scale values of the soft scalar masses after RG evolution.  
By focusing on the mass differences of Eq.~(\ref{stermmass}),
these ambiguities cancel each other out.

  On the other hand, if all the MSSM soft scalar masses are measured,
the low scale values of the soft scalar mass combinations 
\bea 
S &=& Tr(Y\,m^2),\\
S_{B-L} &=& Tr[(B\!-\!L)m^2],\nnmb
\eea
provide useful information about the high scale theory, and can
be used to test possible GUT embeddings of the MSSM~\cite{dgut}.

\subsection*{Flavor Splitting Effects}

  New physics can obscure the underlying flavor structure of
the soft scalar masses.  Family-universal soft masses
derived from a theory containing new physics can generate a 
low-energy spectrum that does not appear to be family-universal after
it is evolved back up to the high scale without including this new
physics.  This is true even if the new physics couples 
in a flavor universal way to the MSSM.  We presented a particular
example of this in Section~\ref{gutmult}, where the new physics
took the form of complete GUT multiplets having no superpotential
couplings with the MSSM sector.

  There are four pairs of soft mass combinations that 
are helpful in this regard~\cite{iblop}.
By comparing these pairs (at any given scale), it is sometimes possible
to obtain clues about the underlying flavor structure of the MSSM 
soft masses.  These combinations are:
\beq
\begin{array}{cclcccl}
m^2_{A_3} &=& 2\,m_{L_3}^2-m_{E_3}^2&\leftrightarrow~~&
m^2_{A_1} &=& 2\,m_{L_1}^2-m_{E_1}^2\\
m^2_{B_3} &=& 2\,m_{Q_3}^2 - m_{U_3}^2 - m_{D_3}^2&\leftrightarrow~~&
m^2_{B_1} &=& 2\,m_{Q_1}^2 - m_{U_1}^2 - m_{D_1}^2\\
m^2_{X_3} &=& {2}\,m_{H_u}^2 - 3\,m_{U_3}^2&\leftrightarrow~~&
m^2_{X_1} &=& {2}\,m_{L_1}^2 - 3\,m_{U_1}^2\\
m^2_{Y_3} &=& {3}\,m_{D_3}^2 + 2\,m_{L_3}^2 - 2\,m_{H_d}^2&\leftrightarrow~~&
m^2_{Y_1} &=& {3}\,m_{D_1}^2
\end{array}
\label{masscombo}
\eeq
If the high scale soft scalar masses are family-universal, 
we expect that each of these pairs, with the possible 
exception of the $m_{X_i}^2$,
to be roughly equal at the low scale in the MSSM.  The $m_{X_i}^2$
combinations are expected to match only if $S=0$ as well.

  To apply the soft mass combinations in Eq.~(\ref{masscombo}),
one should compare them to the splitting between individual
soft masses after running all soft masses up to the high scale
within the MSSM (without new physics).  For instance, an inequality
of the form
\beq
|m_{B_3}^2-m_{B_1}^2| \ll max\left\{|m_{Q_3}^2-m_{Q_1}^2|,\,
|m_{U_3}^2-m_{U_1}^2|,\,|m_{D_3}^2-m_{D_1}^2|\right\},
\eeq
is suggestive of high scale family-universality or a particular
relationship between the $Q$, $U$, and $D$ soft masses
that has been obscured by new physics.  This can arise from
GUT multiplets as in Section~\ref{gutmult}, 
or from a heavy RH neutrino sector as in Section~\ref{neut}.  
Note that heavy neutrinos can disrupt the relationships
between the $A$, $X$, and $Y$ pairs.

\end{document}